\documentclass[arguments]{aastex631}

\usepackage{booktabs}
\usepackage{array,multirow}
\usepackage{tabularx}
\usepackage{rotating}
\shorttitle{Accretion in the MAB}
\shortauthors{Deienno et al.}
\graphicspath{{./}{figures/}}

\begin{document}

\title{Accretion and Uneven Depletion of the Main Asteroid Belt}

\correspondingauthor{Rogerio Deienno}
\email{rogerio.deienno@swri.org, rdeienno@boulder.swri.edu}

\author[0000-0001-6730-7857]{Rogerio Deienno}
\affiliation{Solar System Science \& Exploration Division, Southwest Research Institute, 1050 Walnut Street, Suite 300, Boulder, CO 80302, USA}

\author[0000-0002-4547-4301]{David Nesvorn\'y}
\affiliation{Solar System Science \& Exploration Division, Southwest Research Institute, 1050 Walnut Street, Suite 300, Boulder, CO 80302, USA}

\author[0000-0001-8933-6878]{Matthew S. Clement}
\affiliation{Johns Hopkins APL, 11100 Johns Hopkins Road, Laurel, MD 20723, USA}

\author[0000-0002-1804-7814]{William F. Bottke}
\affiliation{Solar System Science \& Exploration Division, Southwest Research Institute, 1050 Walnut Street, Suite 300, Boulder, CO 80302, USA}

\author[0000-0003-1878-0634]{Andr\'e Izidoro}
\affiliation{Department of Earth, Environmental and Planetary Sciences, MS 126,  Rice University, Houston, TX 77005, USA}

\author[0000-0002-0906-1761]{Kevin J. Walsh}
\affiliation{Solar System Science \& Exploration Division, Southwest Research Institute, 1050 Walnut Street, Suite 300, Boulder, CO 80302, USA}



\begin{abstract}
The main asteroid belt (MAB) is known to be primarily composed of objects from two distinct taxonomic classes, generically defined here as S- and C-complex. The former probably originated from the inner solar system (interior to Jupiter’s orbit), while the latter probably from the outer solar system. Following this definition, (4) Vesta, a V-type residing in the inner MAB (a $<$ 2.5 au), is the sole D $>$ 500 km object akin to S-complex that potentially formed in-situ. This provides a useful constraint on the number of D $>$ 500 km bodies that could have formed, or grown, within the primordial MAB. In this work we numerically simulate the accretion of objects in the MAB region during the time when gas in the protoplanetary disk still existed, while assuming different MAB primordial masses. We then accounted for the depletion of that population happening after gas disk dispersal. In our analysis, we subdivided the MAB into five sub-regions and showed that the depletion factor varies throughout the MAB. This results in uneven radial- and size-dependent depletion of the MAB. We show that the MAB primordial mass has to be $\lesssim$ 2.14$\times$10$^{-3}$ Earth masses. Larger primordial masses would lead to the accretion of tens-to-thousands of S-complex objects with D $>$ 500 km in the MAB. Such large objects would survive depletion even in the outer sub-regions (a $>$ 2.5 au), thus being inconsistent with observations. Our results also indicate that S-complex objects with D $>$ 200-300 km, including (4) Vesta, are likely to be terrestrial planetesimals implanted into the MAB rather than formed in-situ.
\end{abstract}



\section{Introduction} \label{sec:intro}

The main asteroid belt (MAB), situated between the orbits of Mars and Jupiter, contains over 1 million objects with diameter $D$ larger than $\approx$1 km \citep{Bottke2020} .  It also has a  cumulative mass of only about 1/2,000 that of Earth \citep[$\approx$5$\times$10$^{-4}$ M$_{\oplus}$;][where M$_{\oplus}$ represents one Earth mass]{DeMeo2013,DeMeo2014}. The reason  why the MAB has such a tiny total mass has been, for years, a subject of  debate in the literature \citep[e.g.,][]{wetherill80,chambers01,raymondetla06,Walsh2011,Izidoro2015,Roig2015,Deienno2016,Deienno2018,Nesvorny2017,Clement2019,Deienno2022}. 

The MAB can be characterized using two broad taxonomic groupings of asteroids, S- and C-complex \citep{Gradie1982,Mothe-Diniz2003,DeMeo2014}. Spectroscopic analysis \citep{Burbine2002} shows the S-complex to be relatively dry, with less than 0.1\% water by mass, and many constituents are related to inner solar system ordinary chondrites \citep{Robert1977,Che2023}. Many C-complex bodies, on the other hand, potentially carry about 5–20\% water by mass content and seem more akin to outer solar system carbonaceous chondrites \citep{Kerridge1985,Che2023}. The radial distance from the star where the division between inner and outer solar system occurred early in solar system history is poorly defined. The mechanism responsible for diving the two primordial solar system reservoirs that sourced these distinct taxonomic classes is also a matter of debate \citep[e.g.][]{Kruijer2017,Kruijer2020,Brasser2020,Lichtenberg2021,Johansen2021,Izidoro2022,Morbidelli2022}. However, there is a consensus in the literature that water/volatile-poor (S-complex) and water/volatile-rich (C-complex) formed in the respectively regions interior and exterior to the orbit of Jupiter \citep[e.g.;][]{Hellmann2023}

Several works in the literature \citep[e.g.,][to cite a few]{Walsh2011,Walsh2012,Roig2015,Izidoro2016,Deienno2016,Deienno2018,Nesvorny2017,Raymond2017a,Raymond2017b,Clement2019,Lykawka2023} have attempted to find dynamical models that can reproduce both the low mass and taxonomic mixing observed in the current MAB. In general, these models largely agree with the above interpretation that the MAB's taxonomic distribution can be reproduced if the S- and C-complex groups are assumed to have formed in regions interior and exterior to the orbit of Jupiter, respectively, and were transported into the MAB at later stages after their formation. The same models, however, disagree  about the amount of primordial mass in solids that populated the MAB prior to the implantation of planetesimals \citep{Raymond2017a,Raymond2017b}. 

There are currently two  different views for explaining the MAB's low mass. On one side, it was proposed that solids in the protoplanetary disk followed a smooth radial distribution of mass \citep{Weidenschilling1977,Hayashi1981}, and as a result the primordial MAB was composed of several Earth masses. In this case, the MAB's primordial mass had to have been severely depleted at the earliest stages of the solar system history. Early depletion seems necessary to reproduce the abundance of highly siderophile elements measured on the crust and mantle of asteroid (4) Vesta \citep{Zhu2021}. 

Early depletion of the MAB (while gas in the protoplanetary disk still existed) could happen in the following ways. Disk-planet tidal interactions \citep[e.g.][]{Nelson2018} could potentially lead for a putative early migration episode for Jupiter and Saturn \citep{Walsh2011}.  Alternatively, convergent migration of growing protoplanets could result in large objects forming exterior to 1.5 au to migrate towards 1 au \citep{Broz2021,Woo2023}. In both cases, most of the pre-existing objects between 1.5 and 4 au would be removed during the gas disk's lifetime, thus, leading to an early depletion of the MAB region. How early is early in these models depends on many disk parameters, with the real answer poorly constrained.

After gas disk dispersal, early MAB depletion \citep[i.e., during terrestrial planet formation;][see also \citet{Nesvorny2021}]{Clement2018} may occur in the following ways. A violent giant planet dynamical instability \citep{Clement2019}\footnote{Often associated with evolution paths where Jupiter and Saturn acquire highly eccentric orbits, and do not necessarily satisfy outer solar system constraints \citep{Nesvorny2012,Deienno2017}.} would lead to about 99.9\% depletion of the MAB primordial mass. Alternatively, Jupiter and Saturn, while locked in 2:1 mean motion resonance with eccentric orbits \citep{Clement2021a,Clement2021b}, would induce chaotic evolution of secular resonances within the MAB region \citep{Izidoro2016}. Both models could lead to the depletion of the MAB in timescales of the order of 5-10 Myr after gas disk dispersal \citep{Clement2018,Lykawka2023}.

The story is much simpler if the primordial MAB was partially or fully devoid of material initially \citep[e.g.][]{Hansen2009,Izidoro2015}.  This scenario is supported by cosmochemical evidence \citep[e.g.,][]{Warren2011,Bollard2017,Kruijer2017,Kruijer2020,Nanne2019,Spitzer2020,Burkhardt2021} suggesting that the solar system's building blocks formed in concentric rings at various radial distances around the Sun \citep{Izidoro2022,Morbidelli2022}, similar to ring-structured protoplanetary disks observed by the Atacama Large Millimeter Array \citep[ALMA; e.g.,][]{Andrews2018,Huang2018}. In this case, the observed total mass of the MAB would result from the implantation of asteroids of S- and C-complex taxonomic classes \citep[see also \citet{Nesvorny2023}]{Raymond2017a,Raymond2017b}, combined with the mass of S-complex primordial asteroids that formed in the MAB region. The total implanted mass (combined with the MAB primordial mass), however, likely never exceeded ten times the current MAB's total mass, as giant planet dynamical instabilities that were well-tested against outer solar system constraints often provide only about 75-90\% overall depletion rates \citep{Roig2015,Deienno2016,Deienno2018,Deienno2022,Nesvorny2017}. Alternatively, models of pebble accretion \citep{Levison2015a,Johansen2015} could also lead to the formation of a low mass MAB. In this scenario, the distinction between where S- and C-complex asteroids would form and their mixing into the MAB region, however, are less clear as all asteroids (planetesimals) would potentially form out of the same pebble flux.

It is clear from the discussion presented in the previous paragraphs that the primordial amount of mass that initially existed in the MAB is a key ingredient to dictate the formation and early evolution of the solar system. Ideally, we would like to account for both collisional and dynamical evolution (and depletion) in a self-consistent way using modern ideas about how the MAB was affected by early solar system processes. The majority of the work discussed so far assumed that (i) mass depletion was driven by dynamical effects, (ii) asteroids were assumed to be massless test particles, and (iii) collisional evolution was not included (i.e., growth and fragmentation). The few works that accounted for collisional evolution of the MAB \citep[e.g.;][]{Bottke2005,Morbidelli2009} mostly concentrated on understanding the evolution of the MAB size-frequency distribution (SFD), but with limited dynamical effects (i.e. assuming MAB collision probabilities and impact velocities were nearly constant through the age of the solar system) or with models that do not reflect present-day thinking on the problem \citep[i.e., assuming that protoplanets formed in the asteroid belt;][]{Bottke2005b}. Given that many of these works did not directly account for dynamical depletion, the amount of primordial mass that may have existed in the MAB is difficult to constrain.  Therefore, to advance in our understanding of not just where but how planetesimals formed and how planets evolved in the early history of the solar system, it is crucial that we understand how much mass existed in the MAB region after planetesimals formed. To do that, it is important that we consider not only the MAB total mass, but also its SFD.

Only three objects in the current MAB have D $>$ 500 km. Those are the asteroids (1) Ceres, (2) Pallas, and (4) Vesta \citep[e.g.,][]{Bottke2005,Bottke2020}. The first two \citep[a B-type and a C-type, respectively;][]{DeMeo2009} are in the C-complex and probably originated from orbits beyond Jupiter \citep[e.g.;][]{Hellmann2023}. In fact, the formation location of asteroid (2) Pallas is poorly understood, but (1) Ceres has recently been proposed to originate beyond Saturn's orbit \citep{Ribeiro2022}. This would help explain the ammonia signatures found on its surface \citep{deSanctis2015}. Asteroid (4) Vesta \citep[a V-type;][]{DeMeo2009} is the sole D $>$ 500 km object that is a member of the S-complex\footnote{Note that in this work we loosely define S-complex asteroids as any non-carbonaceous objects, including (4) Vesta. We will return to this issue and thoroughly discuss the reasons behind our generically defined taxonomic classification and related implications in Section \ref{sec:sfd}, where they will become more relevant.}.  It is therefore associated with objects originating from orbits interior to that of Jupiter. Its existence provides a useful constraint on the number of D $>$ 500 km
bodies that could have formed, or grown, within the primordial MAB. In other words, the primordial mass of the MAB cannot be such that accretion would lead to the formation of too many
S-complex objects with D $>$ 500 km. If this had been the case, it is possible that more than one could survive solar system evolution and would be observed at present days. This is clearly not the case.

In this work, we followed the accretion of planetesimals in the MAB region as a function of their initial total mass and accretion time.  We also considered cases with and without Jupiter in the simulations. Our primordial MAB SFD is based on the findings of \citet{Morbidelli2009} (see Section \ref{sec:model} for discussion). We compare our results with the current number of objects with D $>$ 500 km that, (i) would form during the lifetime of the gaseous solar nebula \citep[3 or 5 Myr; based on the relative ages of ordinary and CB chondrites, respectively;][see also \citet{Weiss2021} regarding paleo-magnetism constrains]{Pape2019,Krot2005}, and (ii) that would survive the aftermath of the giant planet instability \citep{Nesvorny2012,Nesvorny2017,Deienno2017,Deienno2018,Clement2019}, as well as (iii) chaotic diffusion during subsequent solar system evolution \citep[see also \cite{Deienno2016,Deienno2018,Deienno2022}]{Minton2010}. We show that, unless the primordial mass of the MAB was indeed very low ($\lesssim$ 2.14$\times$10$^{-3}$ M$_{\oplus}$)\footnote{Or eventually depleted to this level before accretion starts, i.e. during the first few 100s of kyr after planetesimals formed, while still in the gas-disk phase.} before accretion started, the number of S-complex objects with D $>500$ km that would form and survive solar system evolution largely exceed the number currently observed \citep[i.e., N$_{\rm{S-complex}}$(D $>$ 500 km) = 1; (4) Vesta;][]{Bottke2005,Bottke2020}.

We present our work in the following structure: Section \ref{sec:model} describes our modeling. In Section \ref{sec:dataset} we describe how we have chosen the data set for comparison between model results and the currently observed MAB. Our results are presented in Section \ref{sec:results}, and conclusion in Section \ref{sec:concl}.

\section{Model} \label{sec:model}

\begin{figure}[h!]
    \centering
    \includegraphics[width=0.5\linewidth]{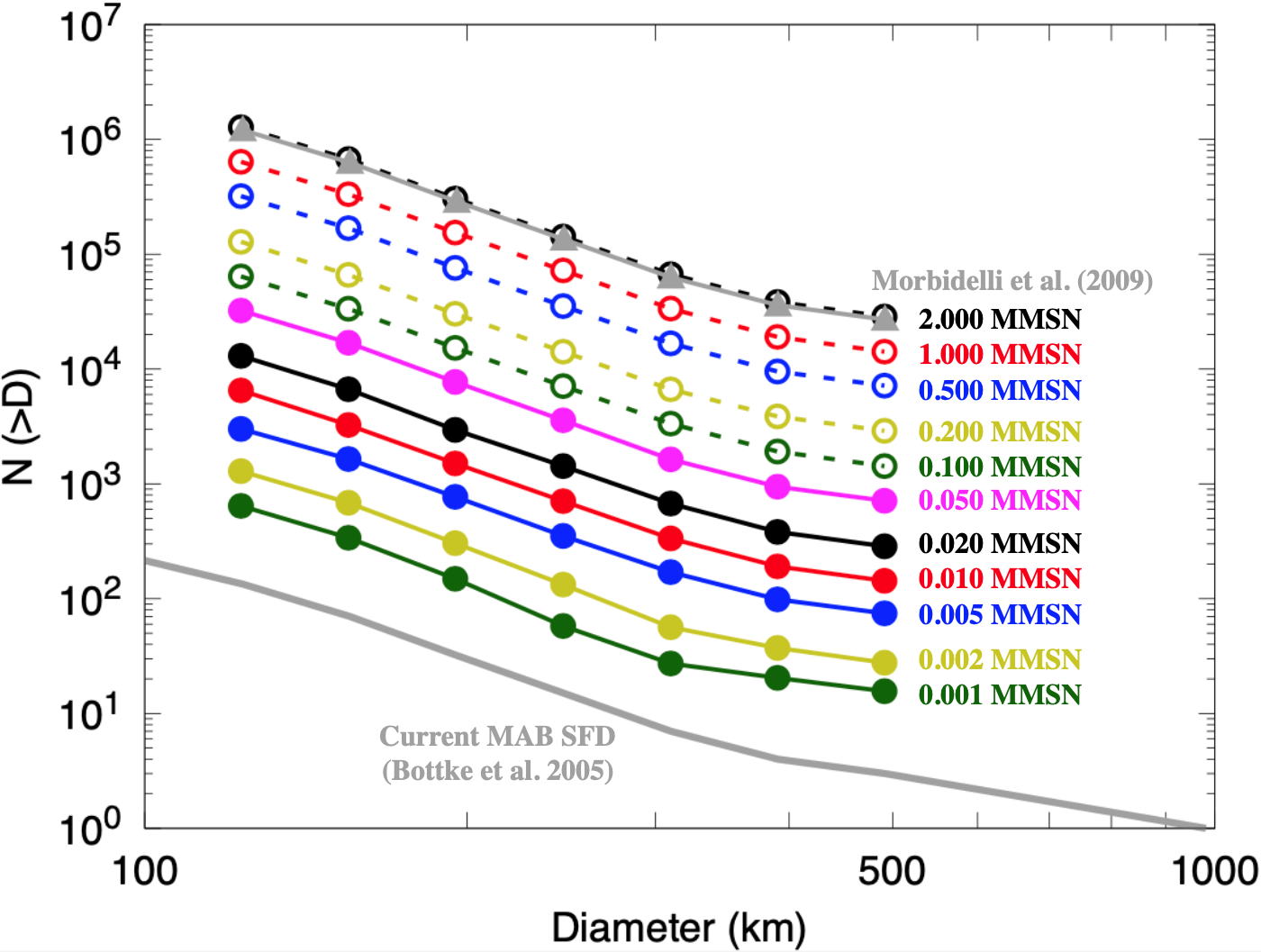}
    \caption{Initial cumulative SFDs adopted for our different primordial MAB populations in terms of MMSN (see main text). For reference we show: (1) Gray line in the bottom representing the current MAB SFD from \citet{Bottke2005}, and (2)  Gray triangles from Figure 7a in \citet{Morbidelli2009}. Other colored symbols/lines are for the different primordial MAB masses considered, in term of MMSN. From top to bottom: 2.000 MMSN (black), 1.000 MMSN (red), 0.500 MMSN (blue), 0.200 MMSN (yellow), 0.100 MMSN (green), 0.050 MMSN (magenta), 0.020 MMSN (black), 0.010 MMSN (red), 0.005 MMSN (blue), 0.002 MMSN (yellow), 0.001 MMSN (green). See main text for the actual mass in Earth masses for each specific case.}
    \label{FigIC}
\end{figure}

To model the evolution of the MAB SFD and the accretion of planetesimals with D $>$ 500 km in the MAB region we first assume that planetesimals were formed within the first 0.5 Myr after Calcium-Aluminum-Inclusions (CAIs) based on the methods by \citet[see also \citet{Lichtenberg2021,Morbidelli2022}]{Izidoro2022}. Therefore, our simulations started at t = 500 kyr.
As for the initial SFD of our planetesimal population we followed \citet[see also \citet{Bottke2005}]{Morbidelli2009}. Although we have some knowledge of the cumulative slope of the SFD for Kuiper belt objects \citep[e.g.,][]{Fraser2014}, not much is known about the SFD of planetesimals that formed in the MAB region \citep[e.g.,][]{Simon2016}. However, we know from \citet{Bottke2005} and \citet{Morbidelli2009} that the current MAB SFD for objects with D $>$ 100 km can only be reproduced if the MAB SFD cumulative slope right after gas disk dispersal was the same as that observed in the current MAB for objects with sizes between 100 km $<$ D $<$ 500 km (Figure \ref{FigIC}). We acknowledge that the primordial MAB SFD may not necessarily be the same as that from the end of the gas disk phase \citep[e.g.,][]{Johansen2015,Simon2016}. Similarly, we acknowledge that different combinations of initial SFD and disruption laws could lead to the conditions reported in \citet[see \citet{Bottke2020}]{Morbidelli2009}. Yet, for simplicity, and to avoid an extensive amount of work dedicated specifically to the combination of initial SFD and disruption laws
that would lead to the finds by \citet{Morbidelli2009}, we decided to assume that the current cumulative slope of the MAB SFD between 100 km $<$ D $<$ 500 km is primordial. We also considered, for consistency, the collision evolution parameters related basalt targets from \citet[i.e., same disruption law considered by \citet{Morbidelli2009}]{Benz1999}. Our results should than be taken with these caveats in mind. We also do not consider pebble accretion in our modeling due to the distinction in the forming regions of S- and C-complex planetesimals that we adopted. Nonetheless, in term of primordial MAB mass, pebble accretion models should likely lead to similar conclusions.

For the total primordial mass of the MAB we assumed that the initial distribution of planetesimals, uniformly distributed between 1.8 and 3.6 au \citep{Deienno2016,Deienno2018}, follows a density profile $\Sigma = \Sigma_0 r^{-\gamma}$, which is analogous to the minimum-mass solar nebula \citep[MMSN;][]{Hayashi1981} radial density profile, and for simplicity we refer to it as such. We chose $\gamma=$ 1  \citep[following values derived from observations of protoplanetary disks, e.g.,][]{Andrews2010}, and a stellar metallicity of 1\% to determine the total mass in solids. We also assumed that only a fraction of the total solid mass was converted into planetesimals \citep[fpl = 0.2, e.g.,][]{Levison2015a,Levison2015b,Johansen2019,Deienno2022}. $\Sigma_0$ in $\Sigma = \Sigma_0 r^{-1}$ is defined as $\zeta \Sigma_{1au}$, where $\Sigma_{1au}$ represents the MMSN surface density at 1 au, equal to 1700 ${\rm g cm^{-2}}$ \citep{Hayashi1981}. We considered five different values for $\Sigma_0$ with $\zeta$ = 0.001, 0.002, 0.005, 0.01, 0.02, 0.05, 0.1, 0.2, 0.5, 1, and 2\footnote{Considering more values of fpl or $\Sigma_0$ is not necessary because values of total disk mass would overlap, e.g., M$_{disk}$ is the same when comparing the pairs (0.1 MMSN; fpl = 0.2) with (0.2 MMSN; fpl = 0.1), or (0.5 MMSN; fpl = 0.1) with (0.1 MMSN; fpl = 0.5), and so on. Therefore, performing simulations with intermediate cases (more values of fpl and $\Sigma_0$) would not necessarily improve or change the results and conclusions.}. This range of parameters results in simulations of MAB primordial masses with initial M$_{disk}\approx$ 0.001, 0.003, 0.007, 0.013, 0.027, 0.068, 0.13, 0.27, 0.68, 1.35, and 2.71 M$_{\oplus}$\footnote{Note that, (1) these latter values ($\ge$ 1 MMSN) are similar to the surface density at $\approx$1 au presumably required to form Earth and Venus \citep[e.g.,][]{chambers01,Walsh2019,Clement2018}, whereas (2) the former values ($<$ 1 MMSN) are consistent with disk models that find planetesimal formation is inefficient in the MAB \citep[e.g.,][]{Drazkowska2018,Lichtenberg2021,Morbidelli2022,Izidoro2022}.}. Planetesimals were initially placed in nearly circular and planar orbits with randomized orbital angles \citep{Deienno2018,Clement2019}. Figure \ref{FigIC} shows the initial cumulative SFDs adopted for our different primordial MAB populations in terms of MMSN.

The exact time and radial distance of Jupiter's formation is unknown \citep[e.g.;][]{Chambers2021,Deienno2022}. We considered cases with and without the gravitational influence of Jupiter in our simulations while following planetesimal accretion during the gas disk phase. The reason for this choice is to test two end-member cases. When considering the effects of Jupiter's gravitational perturbations, we opted to keep Jupiter's semimajor axis fixed near 5.4 au \citep[Jupiter's pre-giant planet instability position is often assumed to be anywhere between 5.4 and 6 au; e.g.,][]{Nesvorny2012,Deienno2017,Clement2018,Lykawka2023}. Not having Jupiter is the same as having it very far away for a long time, whereas having it fixed at 5.4 au is similar to having it form early and/or rapidly migrate to its pre-instability orbit \citep[e.g.,][although still within reasonable distances \citep{Deienno2022}, i.e., with $a\lesssim$ 10--15 au]{Chambers2021}. Lastly, for the simulations where we accounted for Jupiter's perturbations, it is important to acknowledge that, as we cannot self-consistently model hydrodynamical effects of the gas disk acting on Jupiter,  for simplicity, we assumed an initially circular and planar orbit. Additionally, as Saturn is not included in the simulations, we cannot account for any induced effects in the MAB region like chaos arising from the interaction of eccentric Jupiter and Saturn while these worlds reside in their mutual 2:1 mean motion resonance \citep[MMR;][] {Izidoro2016,Lykawka2023}. Still, this approximation may be reasonable, with chaotic effects only expected to become important 5--10 Myr after gas disk dispersal \citep{Izidoro2016}.  We also do not know how long it takes for Jupiter and Saturn to be captured in mutual MMRs and thereby develop significant eccentricities \citep[e.g.,][]{Pierens2011,Pierens2014}. Adding additional complexity to this problem would not necessarily improve our results, but that is an issue that can be more thoroughly tested in future works.

 We followed the accretion of objects within our MAB region for 5 Myr. This interval is presumably the time the gas in the solar nebula dispersed in the outer solar system based on the relative ages of the youngest CB-chondrites to CAIs \citep{Krot2005}. As chondrules may only form and get incorporated into planetesimals while gas still existed in the solar nebula \citep{Johnson2016}, the date of the youngest (latest formed) chondrules are possibly representative of when gas in the protoplanetary nebula may have dispersed for the zone where those chondrules formed. Those ages are also in good agreement with paleo-magnetism constrains, suggesting that the nebular gas dispersal time in the inner and outer solar system was approximately 1.22 Myr $<$ T$_{\rm inner}^{\rm gas}$ $<$ 3.94 Myr and 2.51 Myr $<$ T$_{\rm outer}^{\rm gas}$ $<$ 4.89 Myr \citep{Weiss2021}. We also consider that the gas in the solar nebula dispersed following an exponential timescale \citep[i.e., gas decays as $\exp(-t/\tau_{gas})$;][]{Haisch2001}. We assumed $\tau_{gas}=$ 2 Myr \citep[the average timescale commonly assumed in planet formation and early solar system evolution theories, e.g.,][]{Bitsch2015a,Bitsch2015b,Levison2015a,Levison2015b,Johansen2017,Johansen2019,Johansen2021,Walsh2016,Walsh2019,Broz2021,Deienno2019,Deienno2022,Izidoro2021,Izidoro2022}. Additional tests for $\tau_{gas}=$ 0.5 and 1 Myr produced no substantial differences in our results. For this reason, in Section \ref{sec:results} we only report on simulations with $\tau_{gas}=$ 2 Myr.

To follow the dynamics and accretion of the MAB under the considerations above, we used the code known as LIPAD \citep[Lagrangian Integrator for Planetary Accretion and Dynamics;][]{Levison2012}. As previously described \citep[e.g.,][and references therein]{Deienno2022}, LIPAD is a particle-based (i.e., Lagrangian) code that can follow the collisional/accretional/dynamical evolution of a large number of subkilometer objects throughout the entire growth process to become planets. It uses the concept of tracer particles to represent a large number of small bodies with roughly the same orbit and size \citep[see][for more specific details of the code]{Levison2012}. LIPAD has a prescription for the gaseous nebula from \cite{Hayashi1985}. This gas disk provides aerodynamic drag, eccentricity and inclination damping on planetesimals and planets. The collisional routines from LIPAD use the \cite{Benz1999} disruption laws (see discussion at the begining of this Section). LIPAD is a well-tested code that has been successfully employed in previous studies following the collisional evolution and accretion of centimeter- to kilometer-sized planetesimals on their way to become planets \citep{Kretke2014,Levison2015a,Levison2015b,Walsh2016,Walsh2019,Deienno2019,Deienno2020,Deienno2022,Voelkel2021a,Voelkel2021b,Izidoro2022}, making it ideal for our investigation of the evolution of the MAB SFD while accounting for dynamical effects. 

All planetesimals within all tracer particles are allowed to have self-gravitational interactions and to collide with one-another throughout the simulation. We discarded objects that collisionally evolved to sizes below 1 mm, assuming those would rapidly grind down to dust and no longer contribute to the accretion process \citep{Deienno2020,Deienno2022}.

\section{MAB, Dataset, and Comparison with MPC} \label{sec:dataset}

After we follow accretion in the MAB during the first 5 Myr of the solar system's history, it is necessary to account for the various processes that happened during subsequent 4.5 Gyr evolution after gas disk dispersal. Specifically, the giant planet instability has been found to heavily deplete the MAB \citep{Roig2015,Deienno2016,Deienno2018,Clement2019}. Chaotic diffusion leads to an additional 50 percent of depletion \citep{Minton2010}. Below we gather and reduce numerous works that have simulated the depletion effects of the giant planet instability and map that depletion to different regions in the MAB. This provides an important tool for comparison between formation simulations from this work with observations of the current day MAB.

In this work, we consider the MAB region to be delimited as follows \citep[e.g.,][see also \cite{Roig2015,Nesvorny2017,Clement2019} for similar but slightly different MAB definitions]{Deienno2018}: $i<$ 20$\degr$, $q>$ 1.9 au, $a<$ 3.6 au, and $Q<$ 4.1 au. We want to avoid objects with $q<$ 1.9 au, and $Q>$ 4.1 au; objects on these orbits are likely to have close encounters with Mars and Jupiter, respectively, and their orbits may wander in the $a-e$ and $a-i$ plane. We also do not want any overlap with the Hilda population \citep[semimajor axis centered at $\approx$ 4.1 au; e.g.][]{Roig2015}. We capped the inclination distribution at $i<$ 20$\degr$ as numerical simulations of the orbital evolution of the MAB during the giant planet instability often tend to overpopulate regions with $i>$ 20$\degr$ \citep{Deienno2016,Deienno2018,Nesvorny2017,Clement2019,Clement2020}. These orbital parameter cutoffs allow us to directly compare results with what we may define as the core of the MAB.

\begin{figure}[h!]
    \centering
    \includegraphics[width=\linewidth]{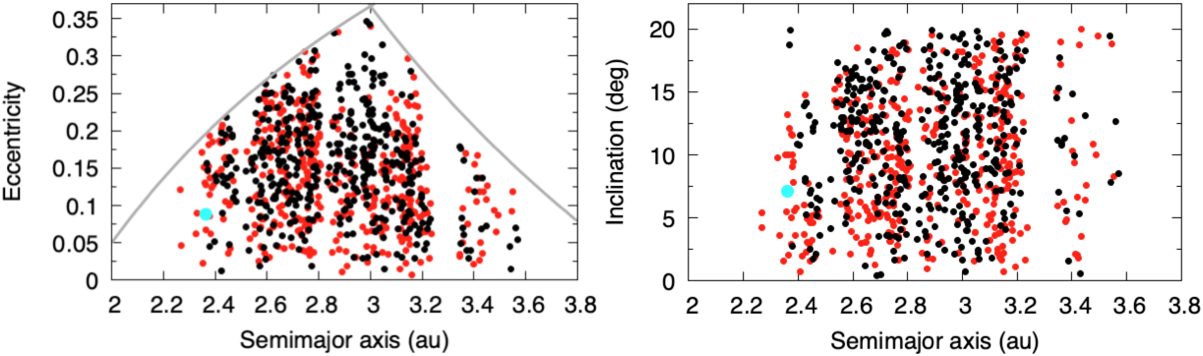}
    \caption{Comparison between the current MAB (see main text for our definition of the MAB) from MPC with $H<$ 9.6 (red) and results from \cite{Deienno2018} (black). Left: Eccentricity $vs$ Semimajor axis. Gray lines demarcate regions where $q=$ 1.9 au and $Q=$ 4.1 au. Right: Inclination $vs$ Semimajor axis. The cyan dot represents the real orbit of asteroid (4) Vesta taken from the MPC data.}
    \label{FigCompare}
\end{figure}

We then use this MAB definition to select objects from the Minor Planet Center (MPC) database\footnote{Data obtained on March 2nd, 2023 at \url{https://minorplanetcenter.net/iau/MPCORB.html}} having $H<$ 9.6, which should correspond to objects with $D\gtrsim$ 50 km (assuming an averaged geometric albedo $\langle p_v \rangle=$ 0.1). We use objects with $D\gtrsim$ 50 km to avoid contamination from main belt fragments likely formed from catastrophic collisions  \citep{Michel2001,Durda2007,Nesvorny2015,Nesvorny2017}. Figure \ref{FigCompare} shows the comparison between real objects taken from MPC and results taken from \cite{Deienno2018} within the previously defined limits of the MAB. Data from \citet[see also \cite{Roig2015,Nesvorny2017,Clement2019,Clement2020} for additional comparison]{Deienno2018} were obtained while modeling the excitation and depletion of the MAB through the giant planet instability \citep{Nesvorny2012,Deienno2017} followed by a subsequent 4.5 Gyr of solar system evolution to account for loss due to chaotic diffusion of long-term unstable orbits \citep{Minton2010}.

Figure \ref{FigCompare} shows very good agreement between results from \cite{Deienno2018} and real data from MPC (compare also the numbers from rows labeled MPC $N_{ast}$ and D18 $N_{ast}^{final}$ from Table \ref{TabCompare}) while accounting for the necessary effects happening during 4.5 Gyr after gas disk dispersal as listed in the beginning of this section. For those reasons, and because results presented in Figure \ref{FigCompare} are derived from simulations that self-consistently satisfied several outer-solar system constraints \citep{Nesvorny2012,Nesvorny2015a,Nesvorny2015b,Deienno2014,Deienno2017}, we refer to data from \cite{Deienno2018} as our fiducial case when estimating the depletion of the MAB due to the giant planet instability followed by 4.5 Gyr of subsequent evolution.

For our depletion analysis, we further divide the MAB definition above into 5 sub-regions: Extended inner Main Belt (EiMB; $a<$ 2.1 au), Inner Main Belt (IMB; 2.1 au $<a<$ 2.5 au), Center Main Belt (CMB; 2.5 au $<a<$ 2.82 au), Outer Main Belt (OMB; 2.82 au $<a<$ 3.25 au), and Extended outer Main Belt (EoMB; $a>$ 3.25 au). The row labeled D18 $D_{fac}$ in Table \ref{TabCompare} shows what the depletion factor is in each of those MAB sub-regions from \cite{Deienno2018}. Table \ref{TabCompare} also reports on the number of asteroids per MAB sub-region in both data from MPC (MPC $N_{ast}$) and from the surviving asteroids from simulations by \citet[D18 $N_{ast}^{final}$ -- values plotted in Figure \ref{FigCompare}]{Deienno2018}. D18 $N_{ast}^{initial}$ represent the number of initial asteroids in the data from \cite{Deienno2018} used for calculating the corresponding depletion factors. The percentage of depletion (D18 $D_{per}$) in each sub-region as well as in the entire MAB is also presented for reference.

\begin{table}[]
    \centering
    \begin{tabular}{lccccc|c}
 \toprule
 \toprule
 & EiMB & IMB & CMB & OMB & EoMB & MAB\\ 
 \cmidrule{2-7}
       MPC $N_{ast}$ & 0 & 43 & 170 & 194 & 24 & 431 \\
\midrule
       D18 $N_{ast}^{final}$ & 0 & 28 & 162 & 221 & 19 & 430 \\
       D18 $N_{ast}^{initial}$ & 1111 & 2222 & 1777 & 2389 & 1944 & 9443 \\
       D18 $D_{fac}$ ($\times$10$^{-2}$) & 0.000 & 1.260 & 9.116 & 9.251 & 0.977 & 4.553 \\
       D18 $D_{per}$ (\%) & 100.000 & 98.740 & 90.884 & 90.749 & 99.023 & 95.447 \\ 
\midrule
       C19$_{all}$ $N_{ast}^{final}$ & 6 & 67 & 90 & 263 & 51 & 477 \\
       C19$_{all}$ $N_{ast}^{initial}$ & 1872 & 7241 & 6526 & 7696 & 7306 & 30641 \\
       C19$_{all}$ $D_{fac}$ ($\times$10$^{-2}$) & 0.320 & 0.925 & 1.379 & 3.417 & 0.698 & 1.557 \\
       C19$_{all}$ $D_{per}$ (\%) & 99.680 & 99.075 & 98.621 & 96.583 & 99.302 & 98.443 \\ 
\midrule
       C19$^{AMD_{\rm JS}}_{P_S/P_J}$ $N_{ast}^{final}$ & 0 & 4 & 28 & 188 & 45 & 265 \\
       C19$_{P_S/P_J}$ $N_{ast}^{initial}$ & 432 & 1671 & 1506 & 1776 & 1686 & 7071 \\
       C19$^{AMD_{\rm JS}}_{P_S/P_J}$ $D_{fac}$ ($\times$10$^{-2}$) & 0.000 & 0.239 & 1.859 & 10.586 & 2.669 & 3.748 \\
       C19$^{AMD_{\rm JS}}_{P_S/P_J}$ $D_{per}$ (\%) & 100.000 & 99.761 & 98.141 & 89.414 & 97.331 & 96.252 \\ 
\bottomrule
\bottomrule
    \end{tabular}
    \caption{Columns represent sub-regions of the MAB (see main text). MPC $N_{ast}$ refers to the current number of asteroids in each MAB sub-region from the MPC data. Simulation data from different works in the literature is provided in sets of four rows. D18 refers to the work by \citet{Deienno2018} \citep[our fiducial case, and similar to the findings from][]{Roig2015,Deienno2016,Nesvorny2017} while C19 refers to the work by \citet{Clement2019} (for a potential  upper limit in depletion). Labels $N_{ast}^{initial}$ and $N_{ast}^{final}$ are for the number of asteroids initially placed in each MAB sub-region and for those whose survived depletion, respectively in each work. $D_{fac}$ represents the depletion factor reported once accounting for the ratio $D_{fac} \approx N_{ast}^{final}/N_{ast}^{initial}$. $D_{fac}$ effectively reports the survival fraction of objects from the simulation. Yet, we prefer to refer to them as a depletion factor as those are the numbers we directly multiply the evolved population in order to account for their depletion. $D_{per}$ is for the percentage of depletion $D_{per} \approx {\rm 100}\times ({\rm 1}-D_{fac})$. Labels $all$ and $^{AMD_{\rm JS}}_{P_S/P_J}$ are for different cuts in the data by \citet[see main text]{Clement2019}.}
    \label{TabCompare}
\end{table}

As previously discussed, the MAB is heavily depleted by the giant planet instability. The level of depletion depends strongly on the number of planet-planet close encounters and the specific orbital evolution of the giant planets, mostly Jupiter, during the giant planet instability. In regards to the MAB evolution and depletion, the most important constraints to satisfy are Jupiter's orbital eccentricity and Jupiter and Saturn's period ratio (P$_S$/P$_J$). Only some giant planet instability simulations, however, are able to satisfy such constraints \citep[see][for a comprehensive review on giant planet instability constraints]{Nesvorny2018a}. Yet, the MAB depletion factor is also highly dependent on the definition of the MAB region. The works by \citet{Roig2015}, \citet{Deienno2016,Deienno2018}, and \citet{Nesvorny2017}, for example, while relying on a small number of simulations that satisfy constraints, report an overall mass depletion about 75\%-90\% for a MAB delimited only by $q>$ 1.9 au and $a<$ 3.2 au. Once further truncating and narrowing the MAB to its core, as shown in Figure \ref{FigCompare} for direct comparison with the MPC catalog, those same giant planet instability simulations lead to an overall depletion of approximately 95.45\% (Table \ref{TabCompare}, D18 $D_{per}$ MAB). Alternatively, the work by \citet{Clement2019} adopted a broader limit for the MAB region between 2 au $\le a \le$ 4 au. That work also relied strongly on the accumulated statistical effects from multiple giant planet instability evolutions without focusing on specific constraints \citep[as done in][]{Roig2015,Deienno2016,Deienno2018,Nesvorny2017}. The resulting overall MAB depletion was reported to be 99.9\%. Although the methodology applied by \citet{Clement2019} is valid, it remains to be confirmed that giant planet instabilities leading up to 99.9\% depletion of the MAB are consistent, not only with constraints related to Jupiter's eccentricity and P$_S$/P$_J$, but also with other solar system constraints obtained from observations of Jupiter Trojans, giant planets regular and irregular satellites, and Kuiper belt objects \citep[which are satisfied in the instability simulation by \citet{Deienno2018}]{Nesvorny2012,Nesvorny2015a,Nesvorny2015b,Deienno2014,Deienno2017}. Yet, for completeness, in this work \citep[see also][]{Deienno2022} we will consider results from \citet{Clement2019} using the following two restrictive definitions (for cases presented in Table \ref{TabCompare}):
\begin{itemize}
    \item C19$_{all}$ refers to results taken from \citet{Clement2019} while excluding runs 4, 7, and 7a\footnote{Cases 4, 7, and 7a (see Figure 2 and Table 2 presented in the work by \citep{Clement2019}) can be ruled out as they result in final orbital excitation and separation of Jupiter and Saturn that are orders of magnitude above the observed values, which in turn lead to 100\% depletion of the entire MAB.}, as well as those with Embryos (runs 1b and 2b)\footnote{As noted by \citet[see also \citet{OBrien2007}]{Clement2019} their simulations struggled in depleting all planetary embryos inserted into the MAB, thus potentially adding unrealistic perturbations that would likely lead to an overestimation of $D_{per}$.}.
    \item C19$^{AMD_{\rm JS}}_{P_S/P_J}$ is also for results from \citet{Clement2019} but while only considering cases where AMD$_{\rm JS}<$ 0.1\footnote{The current angular momentum deficit \citep[AMD;][]{Laskar1997} of Jupiter and Saturn (AMD$_{\rm JS}$) is about 0.0015. Here we consider AMD$_{\rm JS}<$ 0.1 a good proxy because this value is expected to lower during a 100-200 Myr period after the giant planet instability due to dynamical friction exerted by planetesimals interacting with those planets \citep{Nesvorny2012,Deienno2017}.} and 2.3 $\lesssim P_S/P_J \lesssim$ 2.5\footnote{$P_S/P_J\approx$ 2.49 in the current solar system configuration \citep{Nesvorny2012}. In this cut we also avoided the range 2.1 $\lesssim P_S/P_J \lesssim$ 2.3 because in such case Jupiter and Saturn's secular and MMRs would be largely misplaced, thus leading to unseen signatures in the MAB orbital structures \citep{Morbidelli2010,Deienno2016,Deienno2018}.} at the end of the giant planet instability phase (runs 1, 2a, and 5a).
\end{itemize}

The work by \citet{Clement2019} reported on MAB depletion within a few 100s Myr after the end of the giant planet instability, and they did not account for the 50\% additional depletion during subsequent 4.5 Gyr of solar system evolution. For consistency, we divided the asteroid number count by two when calculating values for C19$_{all}$ and C19$^{AMD_{\rm JS}}_{P_S/P_J}$ in each MAB sub-region. These are the final numbers presented in Table \ref{TabCompare}.  We refer to the \citet{Clement2019} depletion factor as C19$_{all}$ $D_{fac}$ and C19$^{AMD_{\rm JS}}_{P_S/P_J}$ $D_{fac}$. The new percentage of depletion from \citet{Clement2019} after these cuts are in much closer agreement with data by \citet[and references therin]{Deienno2018}, with C19$_{all}$ $D_{per} \approx$ 98.44\% and C19$^{AMD_{\rm JS}}_{P_S/P_J}$ $D_{per} \approx$ 96.25\% (not 99.9\% as previously reported at the end of the giant planet instability\footnote{Not accounting for the additional 50\% depletion factor, as presented by \citet{Clement2019}, would lead to C19$_{all}$ $D_{per} \approx$ 96.87\% and C19$^{AMD_{\rm JS}}_{P_S/P_J}$ $D_{per} \approx$ 92.50\%, whose values are then in even closer agreement with \citet{Roig2015,Deienno2016,Deienno2018,Nesvorny2017}.}). C19 cases also present a larger asteroid number difference per MAB sub-region once comparing C19$_{all}$ $N^{final}_{ast}$ and C19$^{AMD_{\rm JS}}_{P_S/P_J}$ $N^{final}_{ast}$ with MPC $N_{ast}$ than D18 $N^{final}_{ast}$. For this reason, and given the larger values for C19$_{all}$ $D_{per}$ and C19$^{AMD_{\rm JS}}_{P_S/P_J}$ $D_{per}$ compared to D18 $D_{per}$, we consider C19$_{all}$ $D_{fac}$ and C19$^{AMD_{\rm JS}}_{P_S/P_J}$ $D_{fac}$ values to be an upper limit for depletion one could expect once accounting for the effects of the giant planet instability and subsequent solar system evolution. 

Our analyses clearly demonstrate that MAB depletion during the giant planet instability is not uniform. The fact that different MAB sub-regions have different depletion factors was first pointed out by \citet{Nesvorny2017}, but never quantified in detail. Table \ref{TabCompare} shows that, regardless of data used, both \citet{Deienno2018} and \citet{Clement2019} report much larger depletion for the EiMB and EoMB than for any other sub-region. In fact, the EiMB is likely to always be fully depleted \citep[e.g.,][]{Bottke2012,Nesvorny2017}. A very large depletion is also expected in the IMB, as the powerful $\nu_6$ secular resonance (resulting from commensurable rates between Saturn's and asteroids' precession frequencies) must cross the IMB region in any formation scenario \citep{Morbidelli2010}.  It is also noticeable that the CMB and OMB present a depletion factor that is about an order of magnitude lower than the IMB. This difference implies that objects that formed in those sub-regions have a much larger chance of surviving in place. This is noteworthy as it implies that these sub-regions are the ones most likely to host asteroids that are native to the MAB. The overall depletion factors listed in the last column of Table \ref{TabCompare} (labeled MAB) can only be used to determine overall mass loss, provided that the initial distribution of solids was even in all MAB sub-regions. The MAB's overall depletion values should not be used in any circumstance to evaluate size dependent depletion, e.g., changes in the SFD. The reason is that objects of different sizes may populate specific MAB sub-regions, thus being more or less likely to be depleted.

\section{Results} \label{sec:results}

We are interested in evaluating both the evolution and depletion of the MAB SFD as well as the number of objects with D $>$ 500 km (4-Vesta-like) that would grow over time within the core of the MAB (see Section \ref{sec:dataset}). Specifically, given the importance of knowing where an object of a certain size is in order to estimate depletion effects (Table \ref{TabCompare}), we are interested in following the growth of these large objects within all 5 MAB sub-regions, i.e., EiMB, IMB, CMB, OMB, and EoMB (see Section \ref{sec:dataset} for definition of such sub-regions). Still, for clarity and simplicity, results reported in Sections \ref{sec:sfdevo}, \ref{sec:sfd} and \ref{sec:depletion} are only presented with respect to the entire MAB, and following our fiducial case (D18; Section \ref{sec:model}, Table \ref{TabCompare}). A detailed breakdown of our results per MAB sub-region while also following C19 models can be found in the Appendix Section.

\subsection{MAB developed SFD}\label{sec:sfdevo}

The current cumulative MAB SFD presents a very characteristic power slope in the 100 km $<$ D $<$ 500 km range. As first pointed by \citet{Bottke2005}, and later confirmed by \citet{Morbidelli2009}, this cumulative SFD slope is a fossil of the MAB population post-gas disk lifetime, and necessary to reproduce the current `bump' observed near D = 100 km \citep[see also Section 4 in][for additional and similar discussion related to C-complex implanted planetesimals into the MAB]{Deienno2022}. For that reason, an important sanity check is to confirm that the cumulative power slope of our evolved SFDs would, at the very least, reasonably resemble the current MAB SFD power slope in the same diameter range by the end of the gas disk lifetime.

\begin{figure}[h!]
    \centering
    \includegraphics[width=0.8\linewidth]{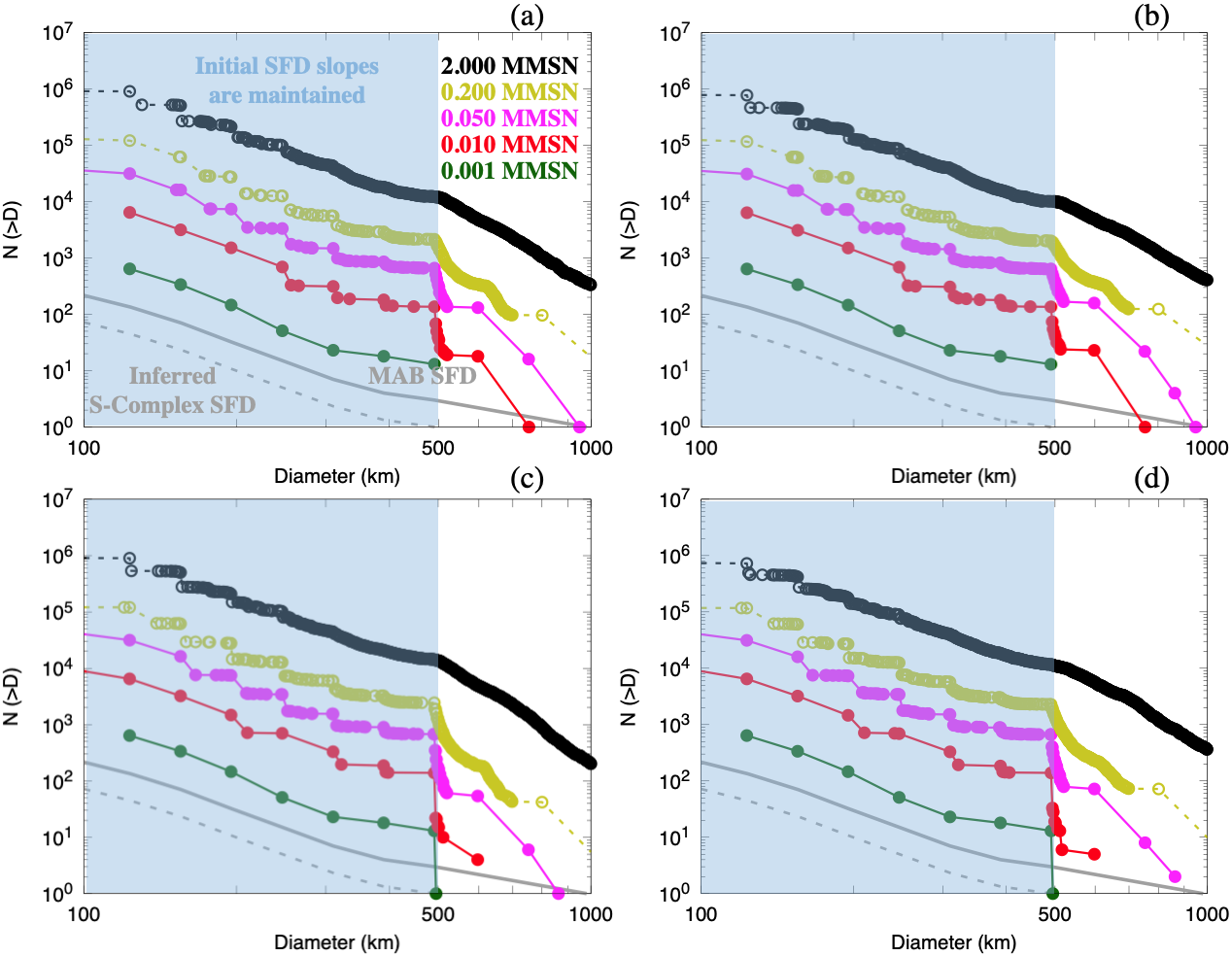}
    \caption{Evolved SFDs. Top panels (a; T = 3 Myr) and (b; T = 5 Myr)   are for simulations where we did not consider Jupiter's perturbation. Bottom panels (c; T = 3 Myr) and (d; T = 5 Myr) are for simulations where we considered Jupiter's perturbation. For reference we show: (1) Solid gray line in the bottom representing the current MAB SFD from \citet{Bottke2005}, and (2)  Dashed gray line our inferred SFD for S-Complex asteroids \citep[see main text.]{Mothe-Diniz2003}. Other colored symbols/lines are the same as in Figure \ref{FigIC} but only a few cases are highlighted to improve visualization. Those are (from top to bottom): 2.000 MMSN (black),  0.200 MMSN (yellow), 0.050 MMSN (magenta), 0.010 MMSN (red), 0.001 MMSN (green). The shaded blue area indicates the target diameter range where the evolved SFDs slope need to resemble that of the current MAB SFD.}
    \label{FigSFDevo}
\end{figure}

In Figure \ref{FigSFDevo} we present the evolved SFDs resulting from our simulations. The resultant SFDs for T = 3 Myr (left panels `a' and `c') and T = 5 Myr (right panels `b' and `d') are very similar. Therefore, having the gas dispersal time be 3 or 5 Myr produces only minor changes \citep[also noted by][]{Walsh2019}, and should not affect the overall growth and evolution of the early MAB. 
Some small differences are observed between simulations where we did not consider Jupiter's influence (top panels `a' and `b') and those where we did (bottom panels `c' and `d'). The main differences are in respect to the number and slope of the distribution for objects with D $>$ 500 km. 
The gravitational perturbation of Jupiter induces larger eccentricities in the MAB. This causes collisions to become more erosive, which frustrates growth \citep[e.g.,][]{Walsh2019}. Yet, of greater importance for this work's analysis is the fact that all evolved SFDs look very similar within the shaded blue region. This means that the influence of Jupiter is not sufficient to change the slope of the MAB SFD for objects with 100 km $<$ D $<$ 500 km.

The fact that our evolved SFDs preserve the primordial power law slope of the cumulative distribution for objects with 100 km $<$ D $<$ 500 km indicates our simulations are compatible with the finds by \citet{Morbidelli2009} and basic characteristics of the current MAB SFD. Thus, we can safely use our evolved SFDs to apply our depletion rates from Table \ref{TabCompare} in order to determine what initial MAB mass (primordial mass that formed at 0.5 Myr after CAIs; Section \ref{sec:model}) can reproduce the current SFD, number of objects with D $>$ 500 km, and total mass observed in S-complex asteroids \citep[dashed gray line in Figure \ref{FigSFDevo} -- under the assumption that the S-complex component of the current MAB represents about 1/4 of the current MAB total mass, i.e., S-complex SFD = 0.25$\times$MAB SFD;][]{Mothe-Diniz2003}. 

\vskip 1.5cm 
\subsection{MAB SFD after uneven depletion} \label{sec:sfd}

As observed in Figure \ref{FigSFDevo}, all evolved SFDs are above the S-complex SFD line by a factor of few to many.
The goal of this section is to determine how the SFDs from Figure \ref{FigSFDevo} would compare with the current S-complex SFD after applying the depletion factors presented in Table \ref{TabCompare}. 
It is important to notice, however, that a very large fraction of the S-complex SFD  as defined in the end of the last Section and presented in Figure \ref{FigSFDevo} is composed by S-type asteroids \citep[recall that we include all non-carbonaceous taxonomic classes, not just S-types, in our generic S-Complex definition, e.g.,][]{DeMeo2014}. 
This is important information because S-type asteroids are believed to have formed from chondrules associated with the parent bodies of ordinary chondrites \citep{Chapman1996,Nakamura2011}, which most likely formed within 2--3 Myr after CAIs \citep[e.g.,][]{Pape2019}.
It may be thus slightly imprecise to directly compare our depleted SFDs with what we generically call the S-complex SFD. As planetesimals in our simulations are assumed to have formed at 0.5 Myr after CAIs \citep{Morbidelli2022,Izidoro2022}, by T = 3 or 5 Myr, they would most likely be differentiated objects \citep[e.g.][]{Lichtenberg2021} like asteroid (4) Vesta. Therefore, for consistency, one may argue that our evolved and depleted SFDs can only be compared to the SFD of differentiated MAB objects. However, it is also important to notice that our SFDs do not significantly evolve from 3 to 5 Myr (Figure \ref{FigSFDevo}). This leads to the conclusion that the late formed (2--3 Myr after CAIs) S-type component of the SFD would only be affected by depletion via giant planet instability, i.e. it would not contribute to accretion. Once again, one may argue that for consistency we should add a second population of asteroids (S-types) to that of our evolved SFDs. We cannot do that, however, because we have zero information of where (in which specific MAB sub-region) large S-type bodies would form. As we will see below, as different MAB sub-regions have different depletion factors (Table \ref{TabCompare}), this makes SFD depletion size-dependent. Thus, assuming that larger late formed S-types would populate one sub-region over the other could lead to very misleading conclusions. Given all that discussion, we argue that the comparison between our results and our generically defined S-complex SFD is valid and correct. Therefore, we can safely assume that the overall evolved SFDs from Figure \ref{FigSFDevo} would not change provided that (i) the population of late formed S-type asteroids was not overwhelmingly large, and (ii) they formed with a cumulative SFD power slope in the range 100 km $<$ D $<$ 500 km similar to what we considered for our primordial population of planetesimals at 0.5 Myr. Both assumptions (i) and (ii) seem necessary to make our results compatible with the current day observed MAB. As we will also show later in this Section, these assumptions will, however, have larger implications for estimating the maximum amount of primordial MAB mass that formed in the MAB at 0.5 Myr after CAIs.

It is important to also say that, in this work, we assume that our SFDs from Figure \ref{FigSFDevo} would not change between the gas disk dispersal time and the onset of the giant planet instability. This is reasonable approximation if we consider that the giant planet instability has to happen anytime within 100 Myr after gas disk dispersal \citep{Nesvorny2018b,Liu2022}, and that collisional timescales can be of the order or larger than that \citep[$t_{coll} \approx \mathcal{O}^{8}$  yrs;][]{Bottke2005}.

\begin{figure}
    \centering
    \includegraphics[width=0.8\linewidth]{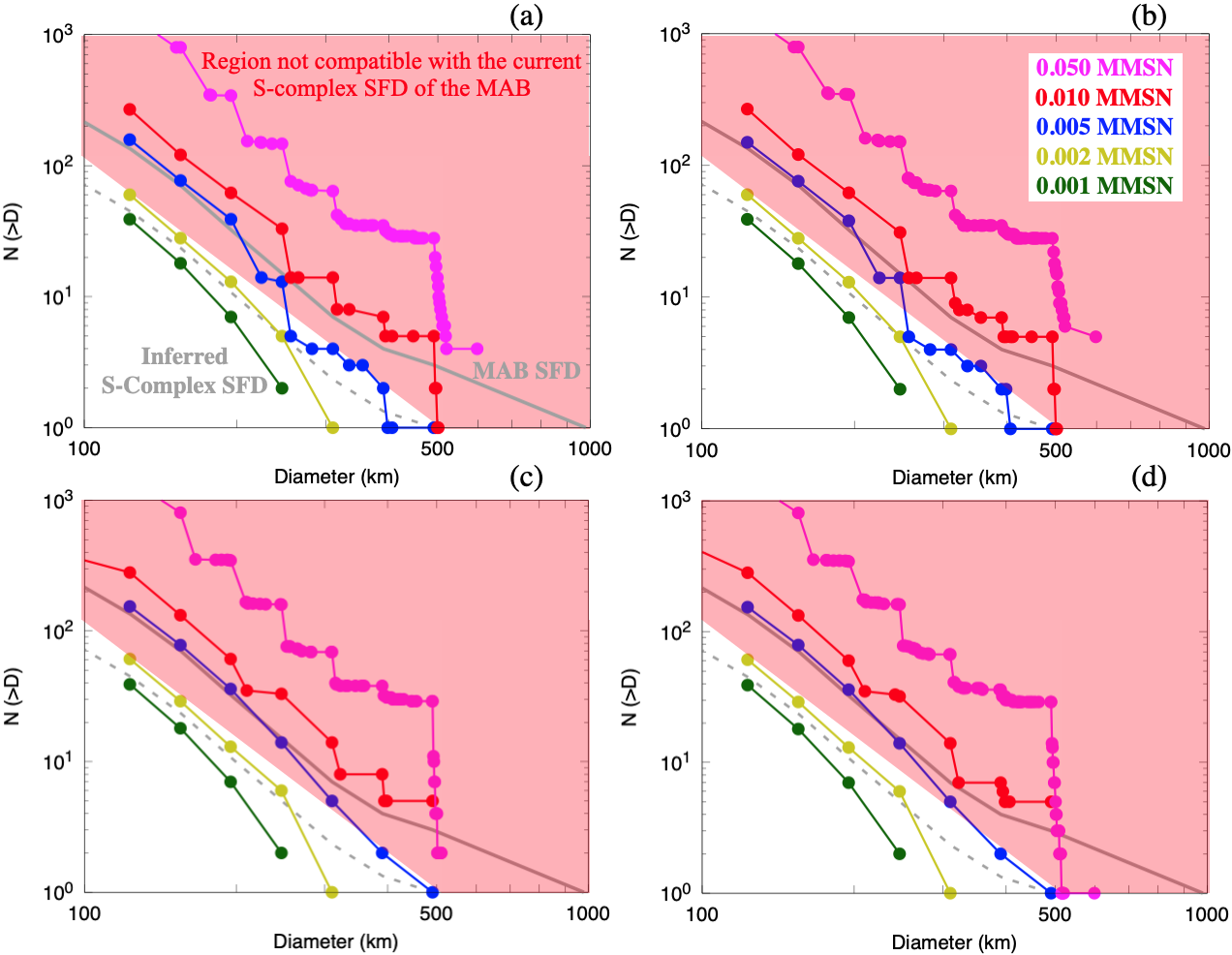}
    \caption{Expected SFDs after depletion following our fiducial parameters (D18 $D_{fac}$) on each individual MAB sub-region. As in Figure \ref{FigSFDevo}, top panels (a; T = 3 Myr) and (b; T = 5 Myr) are for simulations where we did not consider Jupiter's perturbation. Bottom panels (c; T = 3 Myr) and (d; T = 5 Myr) are for simulations where we considered Jupiter's perturbation. For reference we show: (1) Solid gray line in the bottom representing the current MAB SFD from \citet{Bottke2005}, and (2) Dashed gray line our inferred SFD for S-Complex asteroids \citep[see main text in section \ref{sec:sfdevo}.]{Mothe-Diniz2003}. Initial MMSN cases (primordial MAB 
    masses at T = 0.5 Myr) are represented by color and symbols as in Figures \ref{FigIC} and \ref{FigSFDevo} (labeled at top right corner of panel b -- not all cases are shown for clarity purposes):  0.050 MMSN (magenta), 0.010 MMSN (red), 0.005 MMSN (blue), 0.002 MMSN (yellow), and 0.001 MMSN (green). Red shaded area represent depleted SFDs that are not compatible with the current S-complex SFD (gray dashed line), and in most cases comparable or even higher than the total MAB SFD (gray solid line).}
    \label{FigSFDdepletion}
\end{figure}

It is crucial to account for the semi-major axis dependency of depletion when attempting to glean insights into the nature of the primordial MAB from the current MAB SFD.  Figure \ref{FigSFDdepletion} shows results for the expected SFDs post depletion by our fiducial case, D18 $D_{fac}$ (Table \ref{TabCompare}). As we presented in Section \ref{sec:dataset} and Table \ref{TabCompare}, depletion factors are uneven throughout all MAB sub-regions. Therefore,  the estimated post depletion SFDs were calculated by applying D18 $D_{fac}$ to the evolved SFDs from each one of the MAB sub-regions separately\footnote{Following \citet{Deienno2022} we assume depletion caused by the giant planet instability is overall size-independent within each independent MAB sub-region. However, as depletion factors vary in each of the MAB sub-regions, the overall depletion by the giant planet instability may become size-dependent as we illustrate in the main text.}, and combining them after. This is necessary to account for the likelihood of objects being depleted in one MAB sub-region over the other, which in the end, makes depletion size- and semimajor axis-dependent overall. For example, if 200 D $>$ 500 km objects formed in the EiMB and IMB (100 in each sub-region), accounting for D18 $D_{fac}$ for the MAB as a whole (200$\times$0.04553) would lead to about 9 survivors. On the other hand, if we apply the individual D18 $D_{fac}$ values for EiMB and IMB, 100$\times$(0.000+0.0126), the number of D $>$ 500 km survivors to be accounted for in the final MAB SFD would be 1. That would lead for a completely different final SFD and conclusion. This is also the reason for why we previously decided in not accounting for the contribution of late-formed S-types to our evolved SFDs while trying to guess, or assume, which MAB sub-region they would populate. 
As larger objects (D $\gtrsim$ 100 km or so) are always less numerous than smaller km-sized objects, depending on where they formed, uneven depletion may result in a steepening of the depleted SFD when compared to its pre-depletion counterpart (e.g., see post-depletion slopes of yellow and blue SFDs in Figure \ref{FigSFDdepletion} for D $\gtrsim$ 200 km).
The above example clearly demonstrates the importance of considering the depletion factors per MAB sub-region (especially when addressing size dependency) instead of the overall MAB depletion. The latter can only be used as a guide for the overall expected mass depletion. 

 With the post-depletion SFDs evaluated as in Figure \ref{FigSFDdepletion}, we can now use those results to estimate the maximum amount of primordial mass that could have ever existed in the MAB region at T = 0.5 Myr. This maximum estimated primordial mass, however, will be sensitive to our generic definition of the S-complex taxonomy. This is because, as previously discussed in the beginning of this Section, all of the SFDs presented in Figure \ref{FigSFDdepletion} will be a combination of late-formed S-type asteroids and early-formed differentiated bodies. The former, within our definition of the core of the MAB, currently represents a total amount of mass equal to M$_{\rm S-types}$ $\approx$ 3.15$\times$10$^{-5}$ M$_{\oplus}$\footnote{To estimate the total amount of S-type mass we used information from \url{https://mp3c.oca.eu} via `Old Best Values' under the `Search' drop-down menu and following instructions included in `Search by parameters' tab. We restricted our search to semimajor axes between 1.9 au $<a<$ 3.6 au, 0 $<\sin(i)<$ 0.342 ($\approx$ 20$\degr$), and 1.9 au $<q<$ 3.6 au, while selecting `Spectral classes' as {\it S, Sq, Sk, Sr} for S-types only and excluding {\it Q, V, B, C, Cb, Ch, Cg, Cgh, X, Xk, Xc, Xe, E, M, P} \citep[which includes both C-complex and differentiated S-complex bodies;][see also \citet{Delbo2017,Delbo2017db}]{Delbo2019}. We also assumed 3 g/cm$^3$ density for all asteroids and only accounted for those with listed diameters larger than zero. However, we did not use these objects' diameters when comparing with our evolved/depleted SFDs due to the potential effect of data incompleteness.}. Table \ref{TabStypes} presents how this total amount of S-type mass is currently distributed in the MAB by sub-regions (see Current/Post-Depletion row). The following rows in Table \ref{TabStypes}, denoted by Pre-Depletion, show the expected amount of late-formed S-type mass in each MAB sub-region from different models at 2--3 Myr after CAIs. Values presented in row Pre-Depletion by D18, for instance, reflects to our estimate once dividing Current/Post-Depletion values by our fiducial D18 $D_{fac}$ numbers from Table \ref{TabCompare} in each respective MAB sub-region. These values represent the late-formed S-type mass (of a potential second population) that would need to be added to our primordial mass at 2--3 Myr but that would not contribute to subsequent SFD evolution by 3 to 5 Myr. Such amount of mass would only be subject to depletion by the giant planet instability ($D_{fac}$ values in Table \ref{TabCompare}). For each case (i.e., D18, C19$_{all}$, and C19$_{P_S/P_J}^{AMD_{JS}}$), the total amount of late-formed (Pre-Depletion) S-type MAB mass (rightmost column in Table \ref{TabStypes}) was obtained by a simple sum of the numbers in each MAB sub-region. We consider these values to be lower limits in the total S-type mass because $D_{fac}=$ 0 in EiMB (Table \ref{TabCompare}) prevents us from estimating how much S-types could have ever formed in this specific sub-region. 
Therefore, to avoid overestimating the total primordial MAB mass formed at T = 0.5 Myr, the values we considered (Section \ref{sec:model}) in our simulations should be decreased by the amounts shown in the rightmost column of Table \ref{TabStypes}.

\begin{table}[]
    \centering
    \begin{tabular}{l|ccccc|c}
 \toprule
 \toprule
S-type Mass (M$_{\oplus}$) & EiMB & IMB & CMB & OMB & EoMB & MAB\\ 
\midrule
       Current/Post-Depletion  & 0.00 & 7.37$\times$10$^{-6}$ & 1.85$\times$10$^{-5}$ & 5.50$\times$10$^{-6}$ & 9.40$\times$10$^{-8}$ & $\approx$ 3.15$\times$10$^{-5}$ \\
\midrule
\midrule
       Pre-Depletion by D18 & -- & 5.85$\times$10$^{-4}$ & 2.03$\times$10$^{-4}$ & 5.94$\times$10$^{-5}$ & 9.62$\times$10$^{-6}$ & $\gtrsim$ 8.57$\times$10$^{-4}$ \\
\midrule
       Pre-Depletion by C19$_{all}$ & 0.00 & 7.97$\times$10$^{-4}$ & 1.34$\times$10$^{-3}$ & 1.61$\times$10$^{-4}$ & 1.35$\times$10$^{-3}$ & $\approx$ 3.65$\times$10$^{-3}$ \\
\midrule
       Pre-Depletion by C19$_{P_S/P_J}^{AMD_{JS}}$ & -- & 3.08$\times$10$^{-3}$ & 9.95$\times$10$^{-4}$ & 5.15$\times$10$^{-5}$ & 3.52$\times$10$^{-4}$ & $\gtrsim$ 4.48$\times$10$^{-3}$ \\
\bottomrule
\bottomrule
    \end{tabular}
    \caption{Columns represent the estimated amount of mass in each different MAB sub-region  (Section \ref{sec:dataset}). Current/Post-Depletion values where obtained following \citet[see also \citet{Delbo2017,Delbo2017db}]{Delbo2019}. Pre-Depletion values in each MAB individual sub-region where evaluated by dividing Current/Post-Depletion values by D18, D19$_{all}$ and D19$^{AMD_{\rm JS}}_{P_S/P_J}$ $D_{fac}$ from Table \ref{TabCompare}. The total Pre-Depletion MAB S-type mass indicated in the last column of the three bottom lines is a simple sum of the respective columns in the left side. We do not report on Pre-Depletion mass in the EiMB when $D_{fac}=$ 0.}
    \label{TabStypes}
\end{table}

Our results, then indicate that, regardless of Jupiter's inclusion and the exact timing of the nebula gas dispersal, while following our fiducial case, the current overall S-complex SFD (dashed in Figure \ref{FigSFDdepletion}) can only be reproduced if the primordial MAB mass that formed at 0.5 Myr after CAIs was of the order or smaller than about $\approx$ 2.14$\times$10$^{-3}$ M$_{\oplus}$
(i.e., 0.003 M$_{\oplus}$, yellow in Figure \ref{FigSFDdepletion}, decreased by  8.57$\times$10$^{-4}$ M$_{\oplus}$, Table \ref{TabStypes}). A total of 2.14$\times$10$^{-3}$ M$_{\oplus}$ represents about 0.0016 MMSN following our methods in Section \ref{sec:model}, which can still be approximated by 0.002 MMSN (yellow in Figure \ref{FigSFDdepletion}). Similarly, a potential upper limit 
(when considering values from C19$_{all}$ from Tables \ref{TabCompare} and \ref{TabStypes}) on the primordial MAB mass 
should be somewhere between 3.35$\times$10$^{-3}$ M$_{\oplus}$ and 9.35$\times$10$^{-3}$ M$_{\oplus}$ (blue and red in Appendix Figure \ref{FigSFDdepletionC19}, i.e., 0.005 and 0.01 MMSN respectively). Those upper limits in mass were obtained by subtracting 3.65$\times$10$^{-3}$ M$_{\oplus}$ (Table \ref{TabStypes}) from 0.007 and 0.013 M$_{\oplus}$ (Section \ref{sec:model}). Notably, a total of 3.35$\times$10$^{-3}$ M$_{\oplus}$ represents about 0.0025 MMSN, which is very close to our estimate based on our fiducial parameters, whereas 9.35$\times$10$^{-3}$ M$_{\oplus}$ represents about 0.0069 MMSN. However, we should reinforce that those depletion factors were derived by accounting for simulations in which Jupiter and Saturn's AMD ended larger than 0.1 and/or their period ratio P$_S$/P$_J$ exceeded 2.5. When not considering those extreme cases (see Appendix Figure \ref{FigSFDdepletionC19amd} for a comparison), and applying C19$^{AMD_{JS}}_{P_S/P_J}$, our results once again suggest the current S-complex SFD can only be reproduced if the MAB primordial mass at T = 0.5 Myr was of the order or smaller than that represented by 0.002 MMSN (yellow in Appendix Figure \ref{FigSFDdepletionC19amd}). Interestingly enough, however, Table \ref{TabStypes} values associated with C19$^{AMD_{JS}}_{P_S/P_J}$ suggest the late-formed S-type mass at 2--3 Myr should be $\gtrsim$ 4.48$\times$10$^{-3}$ M$_{\oplus}$. Subtracting 4.48$\times$10$^{-3}$ M$_{\oplus}$ of late-formed S-types from a potential primordial population of 0.003 M$_{\oplus}$ (i.e. 0.002 MMSN) that formed at 0.5 Myr would lead to a negative value. This would then imply that according to C19$^{AMD_{JS}}_{P_S/P_J}$ the MAB should have necessarily formed devoid of mass \citep[e.g.][]{Raymond2017b}. Yet, we caution for such conclusion because, although C19$^{AMD_{JS}}_{P_S/P_J}$ case is not as extreme as C19$_{all}$, the final distribution of objects in each MAB sub-region (C19$^{AMD_{JS}}_{P_S/P_J}$ $N_{ast}^{final}$) compared to the MCP catalog (MPC $N_{ast}$; Table \ref{TabCompare}) are not good. In fact, the high estimate of late-formed S-type mass of $\gtrsim$ 4.48$\times$10$^{-3}$ M$_{\oplus}$ is mainly a result of C19$^{AMD_{JS}}_{P_S/P_J}$ $D_{fac}=$ 0.239$\times$10$^{-2}$ in the IMB (with only 4 asteroids surviving compared to 43 from MPC -- a factor $>$10), which leads to a potentially misleading IMB Pre-Depletion mass of $\approx$ 3.08$\times$10$^{-3}$ M$_{\oplus}$. On the other hand, without knowing which sub-region of the MAB those late-formed S-types would populate overall, we cannot take C19$^{AMD_{JS}}_{P_S/P_J}$ on the negative light. If, for example, most of the large estimated mass in the IMB is in few large bodies (i.e. represented solely by the `unknown large diameter end' of the late-formed S-types SFD), even if some primordial mass existed in the IMB, all it is needed is that those few larger bodies are lost. This seems plausible
when considering the differences in depletion rates for different MAB sub-regions (Table \ref{TabCompare}), and it further stresses the importance in considering size- and semimajor axis-dependency as we did in our analyses of Figure \ref{FigSFDdepletion}. Based on all the discussion from this paragraph, we conclude that it is reasonable to be conservative and solely rely on our fiducial estimates based on D18 from \citet{Deienno2018} for the lower limit of $\gtrsim$ 8.57$\times$10$^{-4}$ M$_{\oplus}$ in late-formed S-type mass at around 2--3 Myr. This leads us to the conclusion that the primordial MAB S-complex mass at T = 0.5 Myr after CAIs needs to be $\lesssim$ 2.14$\times$10$^{-3}$ M$_{\oplus}$. 

We should once again stress that our estimate for the total MAB primordial mass (formed at 0.5 Myr after CAIs) above is an upper limit. 
The reason is mostly because, in addition to all that was discussed in the previous paragraph, we cannot evaluate how much extra mass would be added to the S-complex MAB population via implantation during terrestrial planet growth \citep[e.g.,][]{Raymond2017b}. Nonetheless, terrestrial planetesimals would probably be added to the MAB population after gas disk dispersal and would also not contribute to growth in the MAB. Their effective contribution to the shape of the S-complex SFD component is also unclear and we leave this topic for future study. Yet, the fall-off (steepening) of our depleted SFDs for D $\gtrsim$ 200-300 km (Figure \ref{FigSFDdepletion}; e.g. yellow curve) suggest that objects larger than those sizes, including the differentiated asteroid (4) Vesta, were likely terrestrial planetesimals implanted into the MAB \citep[during terrestrial planet accretion;][]{Raymond2017b}, rather than asteroids formed in situ at 0.5 Myr (case of 4 Vesta), or at 2-3 Myr (case of undifferentiated large objects) after CAIs. 

Primordial MAB masses larger than about $\approx$ 2.14$\times$10$^{-3}$ M$_{\oplus}$ formed at 0.5 Myr after CAIs would generate SFDs incompatible with our predicted S-complex SFD, unless depletion occurred much earlier than accretion could have taken place. However, although not shown, an analysis of the temporal evolution of our runs show that the shape of the SFD as presented in Figure \ref{FigSFDevo} for D $>$ 500 km objects develops within the first 0.5 to 1 Myr of evolution (i.e., similar to how we show that the shape of the SFDs for T = 3 and 5 Myr do not change much, we could have shown it for any T $\gtrsim$ 1--2 Myr). This indicates that accretion starts very early in cases where the MAB mass at 0.5 Myr is much larger than $\approx$ 2.14$\times$10$^{-3}$ M$_{\oplus}$. 
Knowing accretion starts very early, we can conclude that our results should be valid regardless of terrestrial planet formation model and/or early evolution of the giant planets \citep[e.g.,][]{Walsh2011,Clement2018,Broz2021,Lykawka2023}. This is even less of a concern in the case of the works by \citet{Clement2018} and \citet{Lykawka2023} as MAB depletion in those scenarios mostly occurs after gas disk dispersal, as we presented in here (i.e., long after accretion happened). Still, our results have larger consequences for those two works as we discuss in the next paragraph.

Our low estimate for the primordial MAB mass points to the conclusion that there is no longer the need for an early giant planet instability to stunt
Mars’ growth \citep{Clement2018}, or for chaotic dynamics to heavily deplete massive disk regions beyond 1--1.5 au \citep{Lykawka2023} after gas disk dispersal. We can confirm the latter. Our results demonstrate that there is no longer the need to heavily deplete a potentially primordial massive disk extending beyond 1--1.5 au. As we showed previouly, such a disk likely never existed, or it would violate MAB constraints (i.e., SFD and number of 4-Vesta-like objects formed; Section \ref{sec:depletion}). We find it difficult, however, from our results alone to explicitly rule out that an early giant planet instability occurring during terrestrial planet formation would not help in stunting Mars. We think
that, depending on the initial profile of the planetesimal disk near 1--1.5 au 
\citep[and whether or not growing embryos in that region would migrate due to interactions with the gas disk component  \citep{Broz2021,Woo2023}]{Izidoro2015,Izidoro2022}, having an early giant planet instability occurring coincident with the later stages of terrestrial planet formation might still be a useful mechanism for preventing Mars from growing too large at 1.5 au \citep{Nesvorny2021}. Nonetheless, our work does reveal the necessity of at the very least revisiting the propositions by \citet{Clement2018} and \citet{Lykawka2023}. This is especially true if considering that most of their simulations that resulted in reasonable terrestrial planets analogs also depleted the MAB to levels that could potentially be incompatible with our new constraints.

\subsection{Growth of D $>$ 500 km objects and their depletion} \label{sec:depletion}

The number of objects with D $>$ 500 km in the MAB, and the particular sub-region of the MAB where they reside, are also important constrains. Knowing how many D $>$ 500 km formed, and more importantly, where did they form and survived depletion can be used as further diagnostic for the maximum amount of mass that could have existed in the primordial MAB. The only known S-complex asteroid with D $>$ 500 km is (4) Vesta, \citep[a V-type;][]{DeMeo2009}, that resides in the IMB. This is important information because very massive primordial MABs could lead to the formation of D $>$ 500 km objects in sub-regions as distant as the OMB, or even EoMB. Depletion factors are uneven, i.e., different in each MAB sub-region, and with higher survival chances in the CMB. That results in high probabilities of D $>$ 500 km objects surviving in sub-regions different than the IMB. In this section we devote our attention to the specific problem of determining how many D $>$ 500 km objects formed in our simulations and in which sub-region of the MAB did they survive depletion.

\begin{figure}
    \centering
    \includegraphics[width=\linewidth]{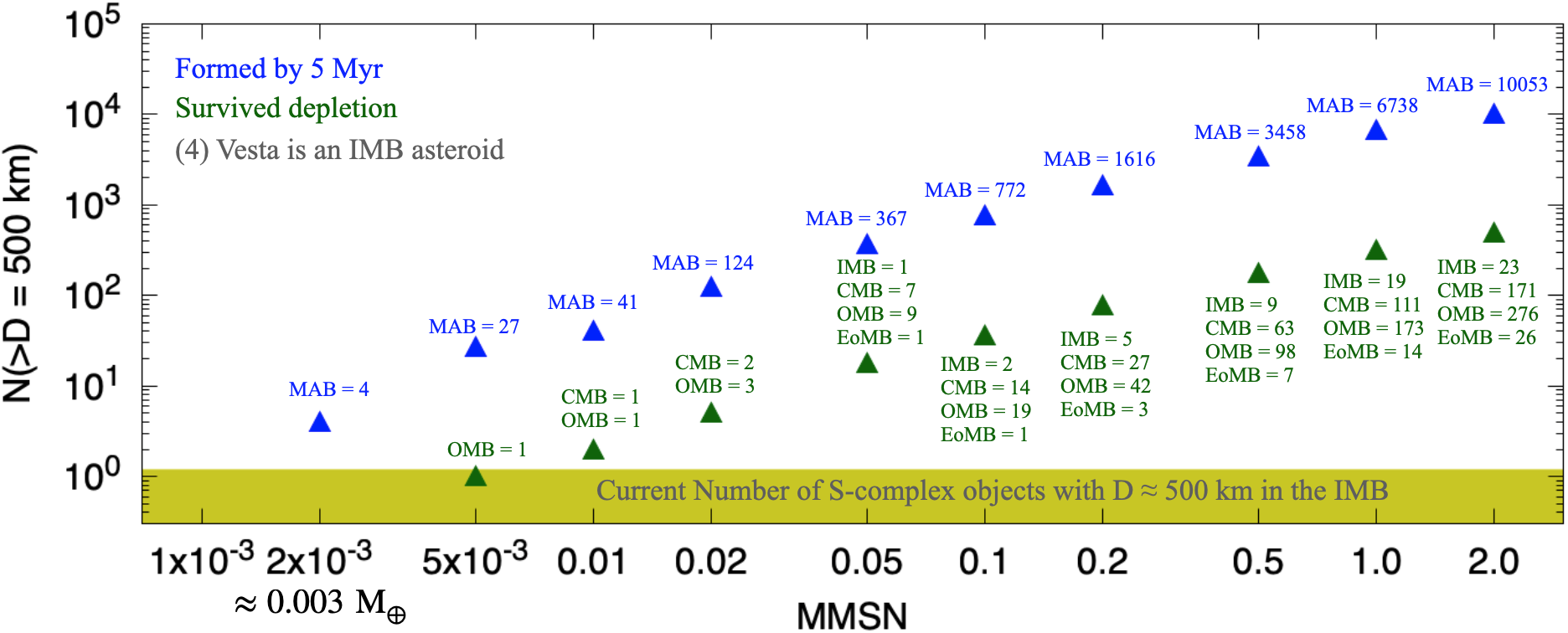}
    \caption{Number of objects with D $>$ 500 km formed within the entire MAB region by T = 5 Myr (blue triangles) as a function of the primordial MAB total mass in terms of MMSN fraction (Section \ref{sec:model}) in a simulation where Jupiter was not present. Blue labels on top of blue triangles show the exact number of objects formed. These numbers are obtained by summing the number of all objects with D $>$ 500 km formed in all 5 MAB sub-regions defined in Section \ref{sec:dataset}, i.e., EiMB, IMB, CMB, OMB, and EoMB. Green triagles refer to the expected number of D $>$ 500 km objects that would survive solar system evolution after gas nebula dispersal when applying our fiducial D18 $D_{fac}$ depletion factors (Table \ref{TabCompare}). Green labels on top of green triangles show how many survived depletion and in which MAB sub-region (see Appendix Table \ref{TabNoJupiter} at T = 5 Myr). Yellow shaded region represents acceptable number of D $>$ 500 km to survive in the IMB after depletion (see main text).}
    \label{FigNVesta}
\end{figure}

Figure \ref{FigNVesta} shows the number of objects with D $>$ 500 km formed within the entire MAB region by T = 5 Myr (blue triangles) as a function of the primordial MAB total mass in terms of MMSN fraction (Section \ref{sec:model}) for a simulation where Jupiter is not present. Figure \ref{FigNVesta} also reports on the expected number of objects after depletion due to effects of the giant planet instability \citep{Nesvorny2012,Deienno2017} and chaotic diffusion over 4.5 Gyr of solar system evolution \citep{Minton2010,Deienno2016,Deienno2018,Deienno2020}, i.e., once applying our fiducial depletion factors from D18 $D_{fac}$ (green triangles) from Table \ref{TabCompare}. For simplicity we do not show figures in this Section with results for the cases where we considered Jupiter or T = 3 Myr. A breakdown of those results, along with those for C19 $D_{fac}$ cases, can be found in the Appendix Tables \ref{TabNoJupiter} and \ref{TabJupiter}. 

The yellow shaded region in Figure \ref{FigNVesta} represents the acceptable number of D $>$ 500 km to survive in the MAB, or more specifically, the IMB after depletion (N $\le$ 1). This number is considered assuming that the initial population of the MAB planetesimals would be of S-complex taxonomic classification (see discussion in Section \ref{sec:intro}). In this case, only asteroid (4) Vesta should be considered as an object of possible formation and survival within the IMB in our study, i.e., N = 1. Still, we report on N $\le$ 1, i.e. N = 0 is a valid result, to account for the possibility where either no objects with D $>$ 500 km formed, or formed but did not survive depletion. In this case, (4) Vesta would not be an object originated in the MAB (or IMB to be more specific; see Figure \ref{FigCompare}) but rather an object that formed in the terrestrial planet region and was later implanted into the IMB during terrestrial planet accretion \citep{Raymond2017b}. We do not consider N = 2 or 3 as an at least marginally successful outcome for two main reasons. First, although the existence of Vesta-like fragments (basaltic objects with V-type taxonomy) that do not show direct orbital link to (4) Vesta and its family suggest the existence of other V-type parent bodies \cite[e.g.,][]{Brasil2017,Troianskyi2023,Burbine2023}, we have no information about their possible origin and size. Therefore, these hypothetical additional V-type parent bodies could likely be much smaller than our D = 500 km size limit. This claim is supported by the absence of a second large and well-defined V-type family in the MAB. Second, we recall that (4) Vesta orbits in the IMB, and as we can see from Figure \ref{FigNVesta}, the formation of a large number of D $>$ 500 km objects would likely lead to their survival in the CMB and OMB. 

An analysis of Figure \ref{FigNVesta}, as well as Appendix Tables \ref{TabNoJupiter} and \ref{TabJupiter}, show that a massive primordial MAB, M$_{disk}$ $\gtrsim$ 0.007--0.013 M$_{\oplus}$ (0.005--0.01 MMSN), leads to the formation of too many D $>$ 500 km objects. As some of them are likely to survive solar system evolution, this would imply that the current MAB should have not only the asteroid (4) Vesta but other larger S-complex asteroids. Furthermore, a very large number of these large objects would reside in sub-regions other than the IMB if the primordial MAB mass was larger than about 0.003 M$_{\oplus}$ (or more specifically 
$\approx$ 2.14$\times$10$^{-3}$ M$_{\oplus}$ -- Section \ref{sec:sfd} -- a direct consequence of uneven depletion). This result strengthens our conclusion from previous Sections in that the MAB primordial mass (at T = 0.5 Myr after CAIs) should be smaller than about 
$\approx$ 2.14$\times$10$^{-3}$ M$_{\oplus}$, or somehow depleted to this level before accretion starts.  

Comparing Appendix Tables \ref{TabNoJupiter} and \ref{TabJupiter} we find that the presence of Jupiter systematically leads to the formation of fewer such large objects, especially within the OMB and EoMB, but also within the CMB. As anticipated in Section \ref{sec:sfdevo}, this is because Jupiter strongly perturbs the sub-regions from OMB to EoMB (and partially the CMB), causing the orbital eccentricity of objects in those sub-regions to increase. A direct  consequence of this eccentricity excitation is that collision velocities among growing objects increase, leading to fragmentation and preventing further growth \citep[e.g.,][]{Walsh2019}. The same effect also explains why depletion seems more aggressive in cases with Jupiter (e.g., compare the number within parentheses in both Appendix Tables \ref{TabNoJupiter} and \ref{TabJupiter} for every case). This is because, when Jupiter is considered, D $>$ 500 km objects are more likely to form only within the EiMB, IMB, and CMB, with the large majority populating the first two sub-regions. Depletion factors associated with the EiMB and IMB are much larger than the ones for the CMB (Table \ref{TabCompare}). This further strengthens our claim that it is very important to account for depletion in each different MAB sub-region separately instead of using the overall depletion factor as a proxy for size-dependent depletion. 

\section{Conclusions} \label{sec:concl}

The primary goal of our work was to determine the maximum possible total mass that could have existed in the primordial MAB region. The current MAB total mass is  tiny \citep[$\approx$5$\times$10$^{-4}$ M$_{\oplus}$;][]{DeMeo2014}. There is a debate in the literature on whether such a small MAB mass was primordial \citep[e.g.;][]{Hansen2009,Levison2015a,Izidoro2015,Izidoro2021,Deienno2016,Deienno2018} or whether the primordial MAB total mass was much larger than currently observed \citep[e.g.;][]{Walsh2011,Clement2019,Lykawka2023}. 

In this work we 
followed accretion (growth and fragmentation) of planetesimals in the MAB during the period when gas still existed in the solar nebula. We assumed different values of total mass for our primordial MAB (see Section \ref{sec:model} for a detailed description of model parameters), and that all of our primordial MAB asteroids would be of inner solar system S-complex taxonomic type \cite[see also discussion in Section \ref{sec:intro}]{DeMeo2014}. We also  attributed a SFD for the MAB primordial population as proposed by previous works that succeeded in reproducing the shape of the current MAB SFD \cite[but with limited dynamical effects]{Bottke2005,Morbidelli2009}. In our work, accretion was followed with the code LIPAD \citep{Levison2012}, meaning that we can in this work self-consistently follow  dynamical and collisional evolution altogether. We studied cases where we consider the gravitational perturbation of Jupiter as well as cases where we did not. The total simulation time was 5 Myr, i.e., presumably the time when gas in the solar nebula dispersed around the MAB region \citep[base on the ages of the yongest Cb-chondrites, as well as on paleo-magnetism constrains;][]{Krot2005,Weiss2021}. Our results confirm previous findings that the power slope of the MAB SFD does not change in the range 100 km $<$ D $<$ 500 km \citep{Bottke2005,Morbidelli2009}, which imply such slope should be primordial. This conclusion is true regardless of considering or not the effects of gravitational perturbation of Jupiter (see Section \ref{sec:sfdevo}).

After following how the MAB SFD evolves under the above circumstances, we considered how the MAB population would be depleted during subsequent solar system evolution. This is important as it is well known that the giant planet instability \citep{Nesvorny2012,Deienno2017}, predicted to have happened at some time after gas disk dispersal, inevitably depletes MAB objects \citep{Roig2015,Deienno2016,Deienno2018,Nesvorny2017,Clement2019}. It is also well known that an additional factor of 2 in depletion is expected due to chaotic diffusion of MAB objects leaking through mean motion resonances over Gyr timescales \citep{Minton2010}. We used results from the literature \citep[e.g.;][]{Deienno2018,Clement2019} to estimate depletion factors that would account for both the giant planet instability and subsequent chaotic diffusion. To do that, we first reduced the data from these works to the same standards. Then, we subdivided the MAB into five sub-regions, i.e. EiMB, IMB, CMB, OMB, and EoMB (see Section \ref{sec:dataset} for a definition of those sub-regions and for a detailed description on MAB limits and data reduction). Finally, we compared our findings with data from real asteroids in the MPC catalog.

The first main result of the present work is that we found the MAB depletion is uneven, i.e., different radial sub-regions of the MAB are depleted at different rates (Table \ref{TabCompare}). This important result was first suggested by \citet{Nesvorny2017}, but never quantified in detail. Here we demonstrate that the CMB and OMB have a much higher survival rate (lower depletion rates) than any other sub-region. This is noteworthy as it implies that these sub-regions are the ones most likely to host asteroids that are native to the MAB. Our results, as shown in Sections \ref{sec:sfd} and \ref{sec:depletion}, clearly demonstrate how such uneven MAB depletion becomes size- and semimajor-axis-dependent. For that reason, we suggest that follow up works use the values for each MAB sub-region derived in Table \ref{TabCompare} separately when accessing depletion of the MAB in a similar way as performed in this work. The overall MAB depletion factor reported in the last column of Table \ref{TabCompare} should only be used as a guide for estimating overall mass depletion, but not size-dependent population depletion, as the latter is also semimajor-axis dependent.

The second main result of the present work is that we found the maximum total mass that could have formed in the primordial MAB at around 0.5 Myr after CAIs is likely to be smaller than $\approx$ 2.14$\times$10$^{-3}$ M$_{\oplus}$. Primordial MAB masses larger than this limit would unavoidably lead to the development of SFDs inconsistent with the current inferred SFD for S-complex asteroids \citep{Mothe-Diniz2003}. Such larger primordial MAB masses would also likely lead to the formation of too many S-complex objects with D $>$ 500 km. Our depletion factors as presented in Table \ref{TabCompare} and following the methods and rational described in Sections \ref{sec:sfd} and \ref{sec:depletion} indicate most objects of this size would survive in the CMB and OMB. Given that the only S-complex object with D $>$ 500 km (4 Vesta) resides in the IMB, we consider such results to be incompatible with observations. Furthermore, we find that objects with D $>$ 200--300 km, including asteroid (4) Vesta, are more likely to be terrestrial planetesimals implanted into the MAB (or more specifically into the IMB) during terrestrial planet growth \citep{Raymond2017b} than asteroids that grew in-situ (see discussion in Sections \ref{sec:sfd} and \ref{sec:depletion} for a detailed rational).

Finally, we conclude by noting that our results are independent of terrestrial planet formation models and/or early evolution of the giant planets \citep[e.g.;][]{Walsh2011,Clement2018,Broz2021,Lykawka2023}. This conclusion is supported by the fact that, for primordial MAB masses larger than $\approx$ 2.14$\times$10$^{-3}$ M$_{\oplus}$, accretion into the MAB starts very early (i.e., within the first few 100s of kyr after planetesimal formation; here considered to be 0.5 Myr after CAIs \citep{Lichtenberg2021,Izidoro2022,Morbidelli2022}). However, most of the MAB depletion happens after gas disk dispersal via a giant planet instability \citep[e.g.;][]{Deienno2018,Clement2019}, or chaotic excitation induced by secular resonance sweeping \citep{Izidoro2016,Lykawka2023}. Therefore, unless depletion took place within the first few 100s of kyr after CAIs, MAB primordial masses larger than $\approx$ 2.14$\times$10$^{-3}$ M$_{\oplus}$ would lead to unobserved features in the current MAB. Such early depletion seems unlikely even in models as those proposed by \citet{Walsh2011} and \citet{Broz2021}.

\section*{Acknowledgements}
We thank the two anonymous reviewers for their positive and constructive comments that helped improving our work. The work of R.D. and K.J.W. was supported by the NASA Emerging Worlds program, grant 80NSSC21K0387. DN's work was supported by the NASA Emerging Worlds program. M.S.C. is supported by NASA Emerging Worlds grant 80NSSC23K0868 and NASA’s CHAMPs team, supported by NASA under Grant No. 80NSSC21K0905 issued through the Interdisciplinary Consortia for Astrobiology Research (ICAR) program. W. F. Bottke's work in this paper was supported by NASA’s Psyche mission through contract NNM16AA09C.


\bibliography{ms}{}
\bibliographystyle{aasjournal}



\appendix

\begin{figure}[h!]
    \centering
    \includegraphics[width=0.6\linewidth]{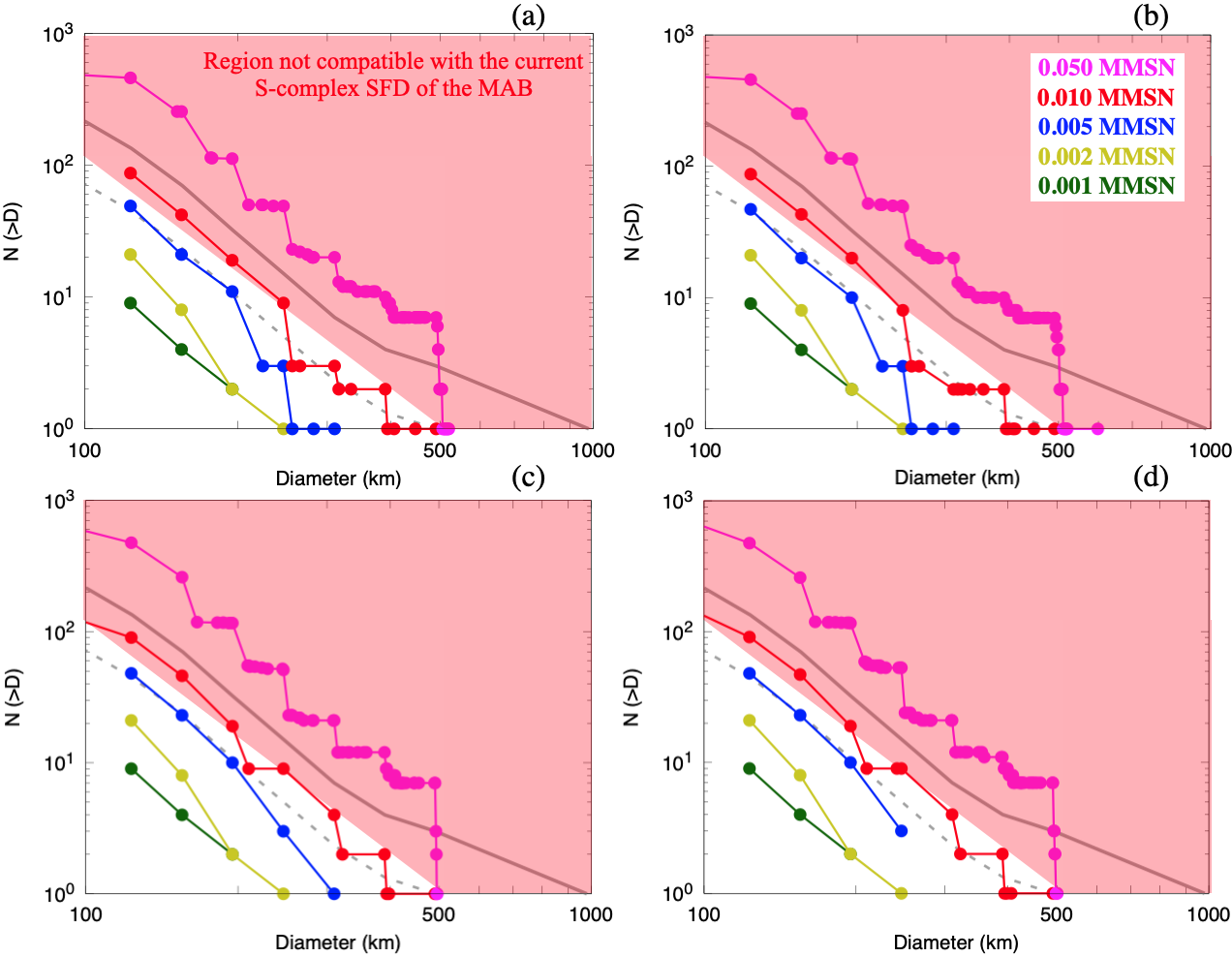}
    \caption{Same as Figure \ref{FigSFDdepletion} but considering the potential upper limit depletion case from C19$_{all}$ from Table \ref{TabCompare}, i.e., where we did not exclude cases in which Jupiter and Saturn's AMD ended larger than 0.1 and/or their period ratio P$_S$/P$_J$ exceed 2.5 (see Figure \ref{FigSFDdepletionC19amd} below for a comparison).}
    \label{FigSFDdepletionC19}
\end{figure}

\begin{figure}[h!]
    \centering
    \includegraphics[width=0.6\linewidth]{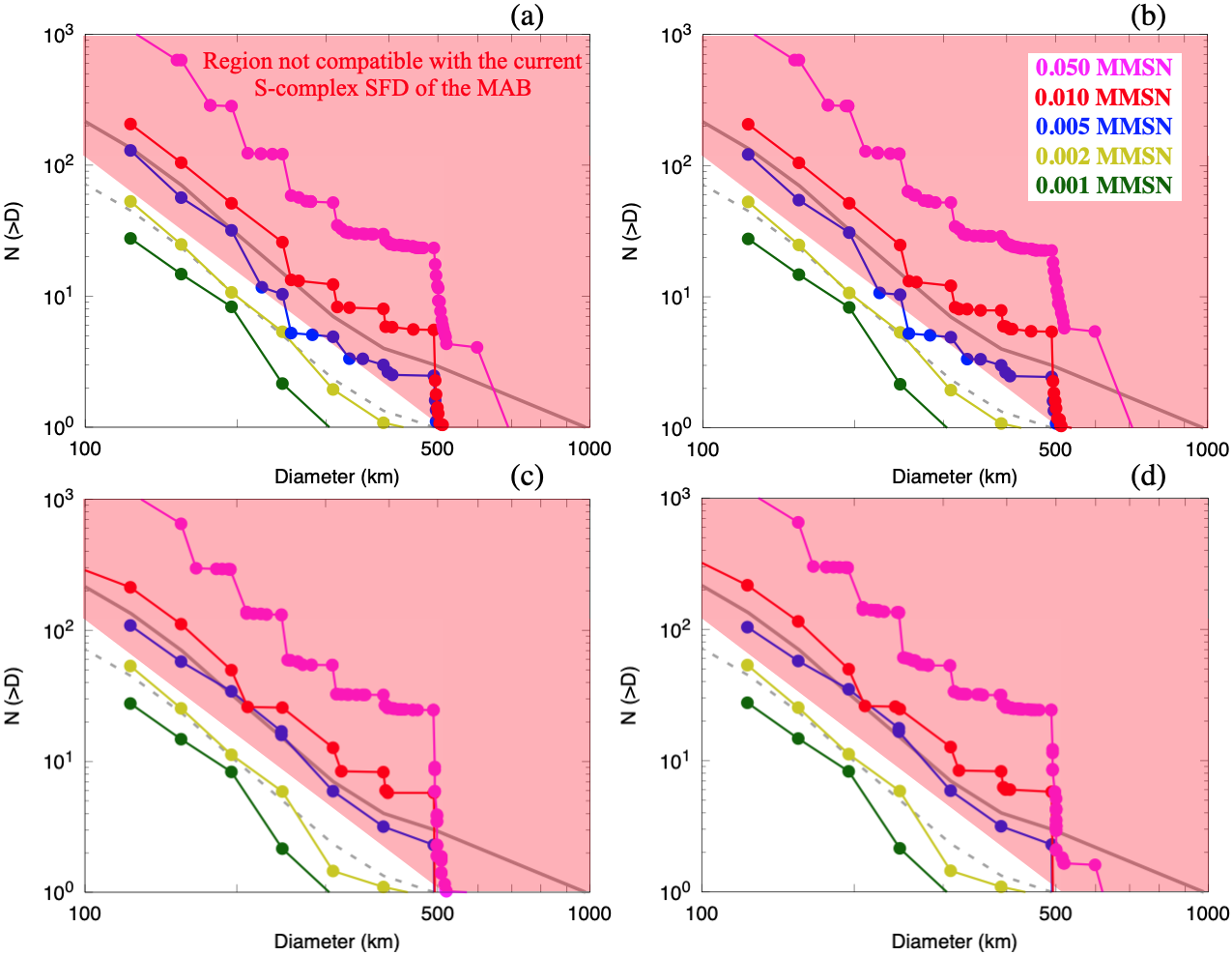}
    \caption{Same as Figures  \ref{FigSFDdepletion} and \ref{FigSFDdepletionC19} while applying depletion factor from C19$^{AMD_{JS}}_{P_S/P_J}$ from Table \ref{TabCompare}, i.e., only considering cases from \citet{Clement2019} where Jupiter and Saturn's AMD remained lower than 0.1 and their period ratio did not exceed 2.5 (similar to cases from \citet[D18 $D_{fac}$ in Table \ref{TabCompare}]{Roig2015,Deienno2016,Deienno2018,Nesvorny2017}).}
    \label{FigSFDdepletionC19amd}
\end{figure}

\begin{table}

   \centering
    \begin{tabular}{c|ccccc|c}
\toprule
\toprule
\multicolumn{7}{c}{Number of D $>$ 500 km objects formed at T = 3 and 5 Myr followed by the expected number of survivors} \\
\multicolumn{7}{c}{after applying ({\bf D18}/C19$_{all}$/C19$^{AMD_{JS}}_{P_S/P_J}$ $D_{fac}$) while {\bf NOT} considering Jupiter} \\
\midrule
\midrule
\multicolumn{7}{c}{{\bf T = 3 Myr}} \\
\midrule
\midrule
MMSN & EiMB & IMB & CMB & OMB & EoMB & MAB \\ 
\midrule

\midrule
\multirow{1}{*}{0.001} &  0 ({\bf 0}/0/0) & 0 ({\bf 0}/0/0) & 0 ({\bf 0}/0/0) & 0 ({\bf 0}/0/0) & 0 ({\bf 0}/0/0) & 0 ({\bf 0}/0/0) \\
\midrule
\multirow{1}{*}{0.002} &  0 ({\bf 0}/0/0) & 2 ({\bf 0}/0/0) & 1 ({\bf 0}/0/0) & 1 ({\bf 0}/0/0) & 0 ({\bf 0}/0/0) & 4 ({\bf 0}/0/0) \\
\midrule
\multirow{1}{*}{0.005} &  4 ({\bf 0}/0/0) & 3 ({\bf 0}/0/0) & 2 ({\bf 0}/0/0) & 8 ({\bf 1}/0/1) & 8 ({\bf 0}/0/0) & 25 ({\bf 1}/0/1) \\
\midrule
\multirow{1}{*}{0.010} &  2 ({\bf 0}/0/0) & 9 ({\bf 0}/0/0) & 11 ({\bf 1}/0/0) & 8 ({\bf 1}/0/1) & 6 ({\bf 0}/0/0) & 36 ({\bf 2}/0/1) \\
\midrule
\multirow{1}{*}{0.020} &  15 ({\bf 0}/0/0) & 27 ({\bf 0}/0/0) & 12 ({\bf 1}/0/0) & 26 ({\bf 2}/1/3) & 23 ({\bf 0}/0/1) & 103 ({\bf 3}/1/4) \\
\midrule
\multirow{1}{*}{0.050} &  42 ({\bf 0}/0/0) & 65 ({\bf 1}/1/0) & 63 ({\bf 6}/1/1) & 83 ({\bf 8}/3/9) & 55 ({\bf 1}/0/1) & 308 ({\bf 16}/5/11) \\
\midrule
\multirow{1}{*}{0.100} &  95 ({\bf 0}/0/0) & 169 ({\bf 2}/2/0) & 128 ({\bf 12}/2/2) & 173 ({\bf 16}/6/18) & 118 ({\bf 1}/1/3) & 683 ({\bf 31}/11/23) \\
\midrule
\multirow{1}{*}{0.200} &  174 ({\bf 0}/1/0) & 395 ({\bf 5}/4/1) & 279 ({\bf 25}/4/5) & 399 ({\bf 37}/14/42) & 263 ({\bf 3}/2/7) & 1510 ({\bf 70}/25/55) \\
\midrule
\multirow{1}{*}{0.500} &  378 ({\bf 0}/1/0) & 746 ({\bf 9}/7/2) & 811 ({\bf 74}/11/15) & 1101 ({\bf 102}/38/117) & 747 ({\bf 7}/5/20) & 3783 ({\bf 192}/62/154) \\
\midrule
\multirow{1}{*}{1.000} &  723 ({\bf 0}/2/0) & 1609 ({\bf 20}/15/4) & 1485 ({\bf 135}/20/28) & 1870 ({\bf 173}/64/198) & 1639 ({\bf 16}/11/44) & 7326 ({\bf 344}/112/274) \\
\midrule
\multirow{1}{*}{2.000} &  865 ({\bf 0}/3/0) & 2244 ({\bf 28}/21/5) & 2101 ({\bf 192}/29/39) & 3528 ({\bf 326}/121/373) & 3073 ({\bf 30}/21/82) & 11811 ({\bf 576}/195/499) \\
\midrule

\midrule
\multicolumn{7}{c}{{\bf T = 5 Myr}} \\

\midrule
\multirow{1}{*}{0.001} &  0 ({\bf 0}/0/0) & 0 ({\bf 0}/0/0) & 0 ({\bf 0}/0/0) & 0 ({\bf 0}/0/0) & 0 ({\bf 0}/0/0) & 0 ({\bf 0}/0/0) \\
\midrule
\multirow{1}{*}{0.002} &  0 ({\bf 0}/0/0) & 2 ({\bf 0}/0/0) & 1 ({\bf 0}/0/0) & 1 ({\bf 0}/0/0) & 0 ({\bf 0}/0/0) & 4 ({\bf 0}/0/0) \\
\midrule
\multirow{1}{*}{0.005} &  5 ({\bf 0}/0/0) & 3 ({\bf 0}/0/0) & 3 ({\bf 0}/0/0) & 8 ({\bf 1}/0/1) & 8 ({\bf 0}/0/0) & 27 ({\bf 1}/0/1) \\
\midrule
\multirow{1}{*}{0.010} &  3 ({\bf 0}/0/0) & 11 ({\bf 0}/0/0) & 11 ({\bf 1}/0/0) & 9 ({\bf 1}/0/1) & 7 ({\bf 0}/0/0) & 41 ({\bf 2}/0/1) \\
\midrule
\multirow{1}{*}{0.020} &  14 ({\bf 0}/0/0) & 31 ({\bf 0}/0/0) & 17 ({\bf 2}/0/0) & 35 ({\bf 3}/1/4) & 27 ({\bf 0}/0/1) & 124 ({\bf 5}/1/5) \\
\midrule
\multirow{1}{*}{0.050} &  46 ({\bf 0}/0/0) & 88 ({\bf 1}/1/0) & 74 ({\bf 7}/1/1) & 92 ({\bf 9}/3/10) & 67 ({\bf 1}/0/2) & 367 ({\bf 18}/5/13) \\
\midrule
\multirow{1}{*}{0.100} &  95 ({\bf 0}/0/0) & 196 ({\bf 2}/2/0) & 151 ({\bf 14}/2/3) & 200 ({\bf 19}/7/21) & 130 ({\bf 1}/1/3) & 772 ({\bf 36}/12/27) \\
\midrule
\multirow{1}{*}{0.200} &  178 ({\bf 0}/1/0) & 404 ({\bf 5}/4/1) & 291 ({\bf 27}/4/5) & 450 ({\bf 42}/15/48) & 293 ({\bf 3}/2/8) & 1616 ({\bf 77}/26/62) \\
\midrule
\multirow{1}{*}{0.500} &  327 ({\bf 0}/1/0) & 708 ({\bf 9}/7/2) & 692 ({\bf 63}/10/13) & 1059 ({\bf 98}/36/112) & 672 ({\bf 7}/5/18) & 3458 ({\bf 177}/59/145) \\
\midrule
\multirow{1}{*}{1.000} &  651 ({\bf 0}/2/0) & 1518 ({\bf 19}/14/4) & 1219 ({\bf 111}/17/23) & 1869 ({\bf 173}/64/198) & 1481 ({\bf 14}/10/40) & 6738 ({\bf 317}/107/265) \\
\midrule
\multirow{1}{*}{2.000} &  660 ({\bf 0}/2/0) & 1861 ({\bf 23}/17/4) & 1881 ({\bf 171}/26/35) & 2982 ({\bf 276}/102/316) & 2669 ({\bf 26}/19/71) & 10053 ({\bf 496}/166/426) \\
\midrule

       \bottomrule
         \bottomrule
    \end{tabular}
    \caption{Detailed information showing the exact number of objects with D $>$ 500 km that formed in each of the individual 5 MAB sub-regions (columns 3--7; number outside parentheses -- see Sections \ref{sec:intro} and \ref{sec:model} for definitions), as well as in the entire MAB (last column; which values were plotted in Figure \ref{FigNVesta} on top of blue triangles at T = 5 Myr). We also present the expected survival of all D $>$ 500 km formed after applying the depletion factors from \citet{Deienno2018} and \citet{Clement2019} using values from Table \ref{TabCompare} in the format ({\bf D18}/C19$_{all}$/C19$^{AMD_{JS}}_{P_S/P_J}$ $D_{fac}$). Boldface values from D18 $D_{fac}$ at T = 5 Myr are those plotted in green in Figure \ref{FigNVesta}. Results are shown for all MMSN (first column) and for the case where Jupiter was {\bf NOT} present in the simulations, at T = 3 (top rows) and 5 Myr \citep[bottom rows; recal that our simulations start at T = 0.5 Myr once we considered planetesimals formed by this time after CAIs;][]{Lichtenberg2021,Izidoro2022,Morbidelli2022}.}
    \label{TabNoJupiter}
\end{table}


\begin{table}

   \centering
    \begin{tabular}{c|ccccc|c}
\toprule
\toprule
\multicolumn{7}{c}{Number of D $>$ 500 km objects formed at T = 3 and 5 Myr followed by the expected number of survivors} \\
\multicolumn{7}{c}{after applying ({\bf D18}/C19$_{all}$/C19$^{AMD_{JS}}_{P_S/P_J}$ $D_{fac}$) when Jupiter {\bf IS} considered in the simulation} \\
\midrule
\midrule
\multicolumn{7}{c}{{\bf T = 3 Myr}} \\
\midrule
\midrule
MMSN & EiMB & IMB & CMB & OMB & EoMB & MAB \\ 
\midrule

\midrule
\multirow{1}{*}{0.001} &  0 ({\bf 0}/0/0) & 0 ({\bf 0}/0/0) & 0 ({\bf 0}/0/0) & 0 ({\bf 0}/0/0) & 0 ({\bf 0}/0/0) & 0 ({\bf 0}/0/0) \\
\midrule
\multirow{1}{*}{0.002} &  0 ({\bf 0}/0/0) & 0 ({\bf 0}/0/0) & 0 ({\bf 0}/0/0) & 1 ({\bf 0}/0/0) & 1 ({\bf 0}/0/0) & 2 ({\bf 0}/0/0) \\
\midrule
\multirow{1}{*}{0.005} &  0 ({\bf 0}/0/0) & 1 ({\bf 0}/0/0) & 1 ({\bf 0}/0/0) & 0 ({\bf 0}/0/0) & 1 ({\bf 0}/0/0) & 3 ({\bf 0}/0/0) \\
\midrule
\multirow{1}{*}{0.010} &  3 ({\bf 0}/0/0) & 10 ({\bf 0}/0/0) & 3 ({\bf 0}/0/0) & 0 ({\bf 0}/0/0) & 0 ({\bf 0}/0/0) & 16 ({\bf 0}/0/0) \\
\midrule
\multirow{1}{*}{0.020} &  11 ({\bf 0}/0/0) & 14 ({\bf 0}/0/0) & 3 ({\bf 0}/0/0) & 4 ({\bf 0}/0/0) & 2 ({\bf 0}/0/0) & 34 ({\bf 0}/0/0) \\
\midrule
\multirow{1}{*}{0.050} &  32 ({\bf 0}/0/0) & 69 ({\bf 1}/1/0) & 24 ({\bf 2}/0/0) & 22 ({\bf 2}/1/2) & 18 ({\bf 0}/0/0) & 165 ({\bf 5}/2/2) \\
\midrule
\multirow{1}{*}{0.100} &  97 ({\bf 0}/0/0) & 153 ({\bf 2}/1/0) & 81 ({\bf 7}/1/2) & 101 ({\bf 9}/3/11) & 47 ({\bf 0}/0/1) & 479 ({\bf 18}/5/14) \\
\midrule
\multirow{1}{*}{0.200} &  195 ({\bf 0}/1/0) & 373 ({\bf 5}/3/1) & 201 ({\bf 18}/3/4) & 246 ({\bf 23}/8/26) & 131 ({\bf 1}/1/3) & 1146 ({\bf 47}/16/34) \\
\midrule
\multirow{1}{*}{0.500} &  404 ({\bf 0}/1/0) & 910 ({\bf 11}/8/2) & 840 ({\bf 77}/12/16) & 1094 ({\bf 101}/37/116) & 587 ({\bf 6}/4/16) & 3835 ({\bf 195}/62/150) \\
\midrule
\multirow{1}{*}{1.000} &  691 ({\bf 0}/2/0) & 1624 ({\bf 20}/15/4) & 1838 ({\bf 168}/25/34) & 2359 ({\bf 218}/81/250) & 1331 ({\bf 13}/9/36) & 7843 ({\bf 419}/132/324) \\
\midrule
\multirow{1}{*}{2.000} &  1069 ({\bf 0}/3/0) & 2319 ({\bf 29}/21/6) & 2629 ({\bf 240}/36/49) & 5084 ({\bf 470}/174/538) & 2514 ({\bf 25}/18/67) & 13615 ({\bf 764}/252/660) \\
\midrule

\midrule
\multicolumn{7}{c}{{\bf T = 5 Myr}} \\

\midrule
\multirow{1}{*}{0.001} &  0 ({\bf 0}/0/0) & 0 ({\bf 0}/0/0) & 0 ({\bf 0}/0/0) & 0 ({\bf 0}/0/0) & 0 ({\bf 0}/0/0) & 0 ({\bf 0}/0/0) \\
\midrule
\multirow{1}{*}{0.002} &  0 ({\bf 0}/0/0) & 0 ({\bf 0}/0/0) & 0 ({\bf 0}/0/0) & 1 ({\bf 0}/0/0) & 1 ({\bf 0}/0/0) & 2 ({\bf 0}/0/0) \\
\midrule
\multirow{1}{*}{0.005} &  0 ({\bf 0}/0/0) & 1 ({\bf 0}/0/0) & 1 ({\bf 0}/0/0) & 0 ({\bf 0}/0/0) & 1 ({\bf 0}/0/0) & 3 ({\bf 0}/0/0) \\
\midrule
\multirow{1}{*}{0.010} &  3 ({\bf 0}/0/0) & 10 ({\bf 0}/0/0) & 3 ({\bf 0}/0/0) & 2 ({\bf 0}/0/0) & 1 ({\bf 0}/0/0) & 19 ({\bf 0}/0/0) \\
\midrule
\multirow{1}{*}{0.020} &  12 ({\bf 0}/0/0) & 22 ({\bf 0}/0/0) & 7 ({\bf 1}/0/0) & 7 ({\bf 1}/0/1) & 2 ({\bf 0}/0/0) & 50 ({\bf 2}/0/1) \\
\midrule
\multirow{1}{*}{0.050} &  48 ({\bf 0}/0/0) & 75 ({\bf 1}/1/0) & 31 ({\bf 3}/0/1) & 35 ({\bf 3}/1/4) & 26 ({\bf 0}/0/1) & 215 ({\bf 7}/2/6) \\
\midrule
\multirow{1}{*}{0.100} &  108 ({\bf 0}/0/0) & 204 ({\bf 3}/2/0) & 98 ({\bf 9}/1/2) & 144 ({\bf 13}/5/15) & 71 ({\bf 1}/0/2) & 625 ({\bf 26}/8/19) \\
\midrule
\multirow{1}{*}{0.200} &  208 ({\bf 0}/1/0) & 379 ({\bf 5}/4/1) & 251 ({\bf 23}/3/5) & 348 ({\bf 32}/12/37) & 193 ({\bf 2}/1/5) & 1379 ({\bf 62}/21/48) \\
\midrule
\multirow{1}{*}{0.500} &  326 ({\bf 0}/1/0) & 815 ({\bf 10}/8/2) & 881 ({\bf 80}/12/16) & 1141 ({\bf 106}/39/121) & 739 ({\bf 7}/5/20) & 3902 ({\bf 203}/65/159) \\
\midrule
\multirow{1}{*}{1.000} &  671 ({\bf 0}/2/0) & 1444 ({\bf 18}/13/3) & 1624 ({\bf 148}/22/30) & 2243 ({\bf 207}/77/237) & 1424 ({\bf 14}/10/38) & 7406 ({\bf 387}/124/308) \\
\midrule
\multirow{1}{*}{2.000} &  805 ({\bf 0}/3/0) & 1981 ({\bf 25}/18/5) & 2202 ({\bf 201}/30/41) & 3847 ({\bf 356}/131/407) & 2212 ({\bf 22}/15/59) & 11047 ({\bf 604}/197/512) \\
\midrule

       \bottomrule
         \bottomrule
    \end{tabular}
    \caption{Detailed information showing the exact number of objects with D $>$ 500 km that formed in each of the individual 5 MAB sub-regions (columns 3--7; number outside parentheses -- see Sections \ref{sec:intro} and \ref{sec:model} for definitions), as well as in the entire MAB (last column). We also present the expected survival of all D $>$ 500 km formed after applying the depletion factors from \citet{Deienno2018} and \citet{Clement2019} using values from Table \ref{TabCompare} in the format ({\bf D18}/C19$_{all}$/C19$^{AMD_{JS}}_{P_S/P_J}$ $D_{fac}$). Results are shown for all MMSN (first column) and for the case when Jupiter {\bf IS} present in the simulations, at T = 3 (top rows) and 5 Myr \citep[bottom rows; recal that our simulations start at T = 0.5 Myr once we considered planetesimals formed by this time after CAIs;][]{Lichtenberg2021,Izidoro2022,Morbidelli2022}.}
    \label{TabJupiter}
\end{table}

\end{document}